\documentclass[twocolumn]{aa}

\usepackage{hyperref}
\usepackage{graphicx}
\usepackage{natbib}
\bibpunct{(}{)}{;}{a}{}{,} 
\usepackage{txfonts}
\usepackage{xcolor}
\usepackage{listings}
\usepackage{acronym}
\usepackage{multirow}
\usepackage{deluxetable}
\usepackage[outdir=./]{epstopdf}

\usepackage[switch]{lineno}

\usepackage{hyperref}
\definecolor{myblue}{HTML}{1F77B4}
\definecolor{mygreen}{HTML}{2CA02C}
\definecolor{myred}{HTML}{D62728}
\definecolor{mymagenta}{HTML}{D33682}
\definecolor{codepurple}{HTML}{C42043}

\hypersetup{
bookmarks=true,         
unicode=true,           
colorlinks=true,        
linkcolor=myred,        
citecolor=myblue,       
filecolor=magenta,      
urlcolor=mygreen        
}

\acrodef{ztf}[ZTF]{Zwicky Transient Facility}
\acrodef{lc}[LC]{light curve}
\acrodef{gp}[GP]{Gaussian Processing}
\acrodef{sn}[SN]{supernova}
\acrodef{sne}[SNe]{supernovae}
\acrodef{obj}[SN\,2020faa]{ZTF20aatqesi}

\newcommand{\swift}{\textit{Swift}}
\newcommand{\galex}{\textit{GALEX}}
\newcommand{\wise}{\textit{WISE}}

\newcommand{\msun}{M$_{\sun}$}

\title{Supernova SN 2020faa - an iPTF14hls look-alike?}
\titlerunning{The iPTF14hls sibling?}

\author{S. Yang\inst{1}
\and 
J. Sollerman\inst{1}
\and{T.-W. Chen}\inst{1}
\and {E. C. Kool}\inst{1}
\and{R. Lunnan}\inst{1}
\and{S. Schulze}\inst{2} 
\and{N. Strotjohann}\inst{2}
\and{A. Horesh}\inst{3}
\and{M. Kasliwal}\inst{4}
\and{T. Kupfer}\inst{5}
\and{A.~A.~Mahabal}\inst{6,7}
\and{F. J. Masci}\inst{8} 
\and{P. Nugent}\inst{9,10}
\and{D. A. Perley}\inst{11}
\and{R. Riddle}\inst{4}
\and{B. Rusholme}\inst{8} 
\and{Y. Sharma}\inst{4}
}
\institute{Department of Astronomy, The Oskar Klein Center, Stockholm University, AlbaNova, 10691 Stockholm, Sweden
\and{Department of Particle Physics and Astrophysics, Weizmann Institute of Science, 234 Herzl St, 76100 Rehovot, Israel}
\and{Racah Institute of Physics, The Hebrew University of Jerusalem, Jerusalem 91904, Israel}
\and{Caltech Optical Observatories, California Institute of Technology, Pasadena, CA 91125, USA}
\and{Texas Tech University, Department of Physics \& Astronomy, Box 41051, 79409, Lubbock, TX, USA}
\and{Division of Physics, Mathematics and Astronomy, California Institute of Technology, Pasadena, CA 91125, USA}
\and{Center for Data Driven Discovery, California Institute of Technology, Pasadena, CA 91125, USA}
\and{IPAC, California Institute of Technology, 1200 E. California, Blvd, Pasadena, CA 91125, USA}
\and{Lawrence Berkeley National Laboratory, 1 Cyclotron Road, Berkeley, CA 94720, USA}
\and{Department of Astronomy, University of California, Berkeley, CA 94720-3411, USA}
\and{Astrophysics Research Institute, Liverpool John Moores University, IC2, Liverpool Science Park, 146 Brownlow Hill, Liverpool L3 5RF, UK}
}

\date{Submitted to A\&A on 2020 September 15}

\begin{document}
\abstract
    {We present observations of \ac{obj}. This Type
      II \ac{sn} displays a luminous \ac{lc} that started to rebrighten from an
      initial decline.
      We investigate this in relation to the famous \ac{sn} iPTF14hls, which received a lot of attention and multiple interpretations in the literature, however whose nature and source of energy still remains unknown.}
    {We demonstrate the great similarity between \ac{obj} and iPTF14hls during the first 6 months, 
      and use this comparison both to forecast the evolution of \ac{obj} and to reflect on the less well observed early evolution of iPTF14hls.} 
    {We present and analyse our observational data, consisting mainly of
      optical \ac{lc}s from the Zwicky Transient Facility in the $gri$ bands as well as a sequence of optical spectra. We construct colour curves, a bolometric \ac{lc}, compare ejecta-velocity and Black-body radius evolutions for the two \ac{sne}, as well as for  more typical Type II \ac{sne}.}
   {The \ac{lc}s show a great similarity with those of iPTF14hls
     over the first 6 months, in luminosity, timescale and colours. Also the
     spectral evolution of \ac{obj} is that of a Type II \ac{sn}, although it probes earlier epochs than those available for iPTF14hls.}
   {The similar \ac{lc} behaviour is suggestive of \ac{obj}
     being a new iPTF14hls. We present these observations now to
     advocate follow-up observations, since most of the more
     striking evolution of \ac{sn} iPTF14hls came later, with \ac{lc}
     undulations and a spectacular longevity. On the other hand, for
     \ac{obj} we have better constraints on the explosion epoch
     than we had for iPTF14hls, and we have been able to
     spectroscopically monitor it from earlier phases than was done
     for the more famous sibling.}
   

\maketitle

\section{Introduction}
\label{section:intro}

The extraordinary supernova (\ac{sn}) iPTF14hls was a Type II \ac{sn},
first reported by \citet[][hereafter A17]{Arcavi17}
as having a long-lived (600+~d) and luminous 
light curve
(\ac{lc})
showing at least five episodes of rebrightening. 
\citet[][hereafter S19]{Sollerman2019} followed the \ac{sn} until 1000 days when it finally faded from visibility.

The spectra of iPTF14hls were similar to those of other hydrogen-rich \ac{sne}, but evolved at a slower pace. 
A17 described a scenario where this could be the explosion of a very massive star that ejected a huge amount of mass prior to explosion. They connect such eruptions with the pulsational pair-instability mechanism.

Following the report of A17, a large number of interpretations were suggested for this unusual object.
These covered a wide range of progenitors and powering mechanisms.
For example, \citet{Chugai18} agreed on the massive ejection scenario, while \cite{AS17} argued for interaction with the circumstellar medium (CSM) as the source for the multiple rebrightenings in the \ac{lc}, which was supported by S19.
\cite{Dessart18} instead suggested a magnetar as the powering mechanism, whereas \cite{Soker18} advocated a common-envelope jet.
\cite{WangWangWang18} proposed a fall-back accretion model for iPTF14hls and \cite{Woosley18} discuss pros and cons of several of the above-mentioned models, and whether the event was indeed a final
explosion.  
\citet{2020MNRAS.491.1384M} interpret the phenomenon as due to a wind from a very massive star.

Taken together, this suite of publications demonstrate how extreme objects like iPTF14hls challenge most theoretical models and force us to expand the frameworks for transient phenomena. 
But iPTF14hls was a single specimen - until now.
 
In this paper, we present observations of \ac{obj},
a Type II \ac{sn} that observationally appears to be similar to iPTF14hls during the first six months. 
We present \ac{lc}s and spectra to highlight this similarity and also add information that was not available for iPTF14hls, like earlier spectroscopy and better constrains on the explosion epoch.
In addition to the ground-based observations, we obtained several epochs of data with the \textit{Neil Gehrels Swift Observatory} \citep[\textit{Swift,}][]{gehrels2004}.
A main aim of this paper is to direct the attention of the community to this active transient, which may - or may not - evolve in the same extraordinary way as did iPTF14hls.
If \ac{obj} become another iPTF14hls, we hope these observations, especially those at the early phases, will be complementary ingredients to iPTF14hls for the community to understand the physics of such peculiar long-lived transients.

The paper is structured as follows.
In Sect.~\ref{sec:data}, we outline the observations and corresponding data reductions, including Sect.~\ref{sec:detection} where we present the detection and classification of \ac{obj}.
The ground-based optical \ac{sn} imaging observations and data reductions are presented in Sect.~\ref{sec:optical},
in Sect.~\ref{sec:space} we describe the \textit{Swift} observations,
a search for a pre-cursor is done in
Sect.~\ref{sec:prediscovery}, the optical spectroscopic follow-up campaign is presented in Sect.~\ref{sec:opticalspectra}, and a discussion of the host galaxy is provided in Sect.~\ref{sec:host}.
An analysis and discussion of the results 
is given in Sect.~\ref{sec:discussion} 
and this is summarised in Sect.~\ref{sec:summary}.

For iPTF14hls,
we follow A17 and adopt a redshift of $z=0.0344$, corresponding to a luminosity distance of 156 Mpc. 
We correct all photometry for Milky Way (MW) extinction, $E(B-V)=0.014$~mag, but make no correction for host-galaxy extinction.
For \ac{obj}, we use $z=0.04106$ (see below), corresponding to a luminosity distance of 187 Mpc (distance modulus 36.36 mag) using the same cosmology as A17. The MW extinction is $E(B-V)=0.022$~mag, and also in this case we adopt no host galaxy extinction.
We follow A17 and use the PTF discovery date as a reference epoch for all phases for iPTF14hls, while for \ac{obj}, we set the first ATLAS detection date as reference epoch.

\section{Observations and Data reduction}
\label{sec:data}

\subsection{Detection and classification}
\label{sec:detection}

The first detection of SN\,2020faa (a.k.a. ZTF20aatqesi) with the Palomar Schmidt 48-inch (P48) Samuel Oschin telescope was on 2020 March 28 ($\mathrm{JD}=2458936.8005$), as part of the \ac{ztf} survey \citep{Bellm_2018,Graham_2019}.
The object had then already been discovered and
reported to the Transient Name Server 
(TNS\footnote{\href{https://wis-tns.weizmann.ac.il}{https://wis-tns.weizmann.ac.il}}) 
by the ATLAS collaboration
\citep{2020TNSTR.869....1T}
with a discovery date of 2020 March 24
($\mathrm{JD_{\rm{discovery}}}=2458933.104$) at $18.28$ mag in the cyan band, and a reported last non-detection
($> 18.57$) 14 days before discovery in the orange band.

The first \ac{ztf} detection was made in the $g$ band,
with a host-subtracted magnitude
of $18.40\pm0.09$~mag, at the J2000.0 coordinates
$\alpha=14^{h}47^{m}09.50^{s}$, $\delta=+72\degr44\arcmin11.5\arcsec$. 
The first $r$-band detection came in 3.6 hours later at $18.50\pm0.10$.
The non-detections from \ac{ztf} include a $g$-band non-detection from 15 days before discovery, 
but this is a shallow global limit ($> 17.46$), whereas the one at 17 days before discovery is deeper at $> 19.37$ mag.
The constraints on the time of explosion for \ac{obj} are thus not
fantastic, but in comparison with the very large uncertainty for iPTF14hls ($\sim100$ days) they are quite useful.

\begin{figure*}
\centering
    \includegraphics[width=0.8\textwidth]{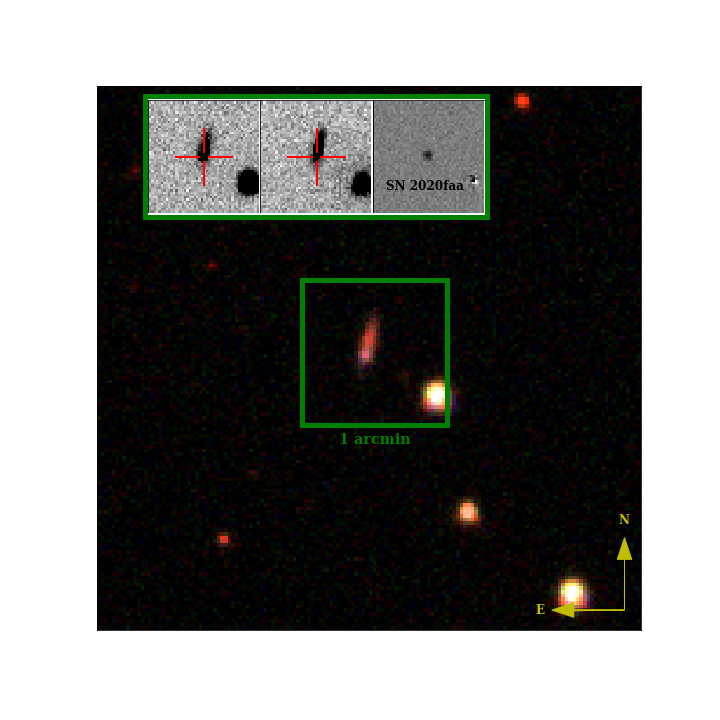}
    \caption{A $gri$-colour composite image of \ac{obj} and its environment, as observed 
    with the P48 telescope on 2020 April 5, eight days after the first \ac{ztf} detection. The $g$-band image subtraction is shown in the top panel.}
    \label{fig:det}
\end{figure*}

\ac{obj} is positioned on the edge on spiral galaxy
WISEA J144709.05+724415.5 which did not have a reported redshift in
the NED\footnote{\href{https://ned.ipac.caltech.edu}{https://ned.ipac.caltech.edu}} catalog, although the CLU catalog \citep{CLU} has it listed as CLU J144709.1+724414 at the same redshift as our spectroscopy provides below. The \ac{sn} together with the host galaxy and the field of view is shown in Fig.~\ref{fig:det}.

\ac{obj} was classified as a Type II \ac{sn}
\citep{2020TNSCR.987....1P}
based on a spectrum obtained on 2020 April 6
with the Liverpool telescope (LT) equipped with the SPRAT (SPectrograph for the Rapid Acquisition of Transients) spectrograph. 
That spectrum revealed broad H$\alpha$ and H$\beta$ in
emission, the blue edge being shifted by $\sim 9000$~km~s$^{-1}$ with respect to the narrow emission line from the galaxy that provided the redshift $z=0.041$ consistent with CLU as mentioned above.
The LT spectrum confirmed the tentative redshift and classification deduced from our first spectrum, obtained with the Palomar 60-inch telescope (P60; \citealp{2006PASP..118.1396C}) equipped with the Spectral Energy Distribution Machine (SEDM;
\citealp{SEDM}). 
That first spectrum was taken already on March 31, but the quality was not good enough to warrant a
secure classification.

\subsection{Optical photometry}
\label{sec:optical}

Following the discovery, we obtained regular follow-up photometry
during the slowly declining phase in $g$, $r$ and $i$ bands
with the \ac{ztf} camera \citep{dekany2020} on the P48. This first decline
lasted for $\sim50$ days, and no further attention was given to the \ac{sn}
during this time.

Later on, after rebrightening started, we also obtained a few epochs
of triggered photometry in $gri$ with the SEDM on the P60.
The \ac{lc}s from the P48
come from the \ac{ztf} pipeline \citep{2019PASP..131a8003M}.
Photometry from the P60 were produced with the image-subtraction pipeline described in \cite{fremling16}, with template images from the Sloan Digital Sky Survey (SDSS; \citealp{ahn14}). 
This pipeline produces PSF magnitudes, calibrated against SDSS stars in the field. 
All magnitudes are reported in the AB system.

\begin{figure*}
\centering
    \includegraphics[width=0.8\textwidth]{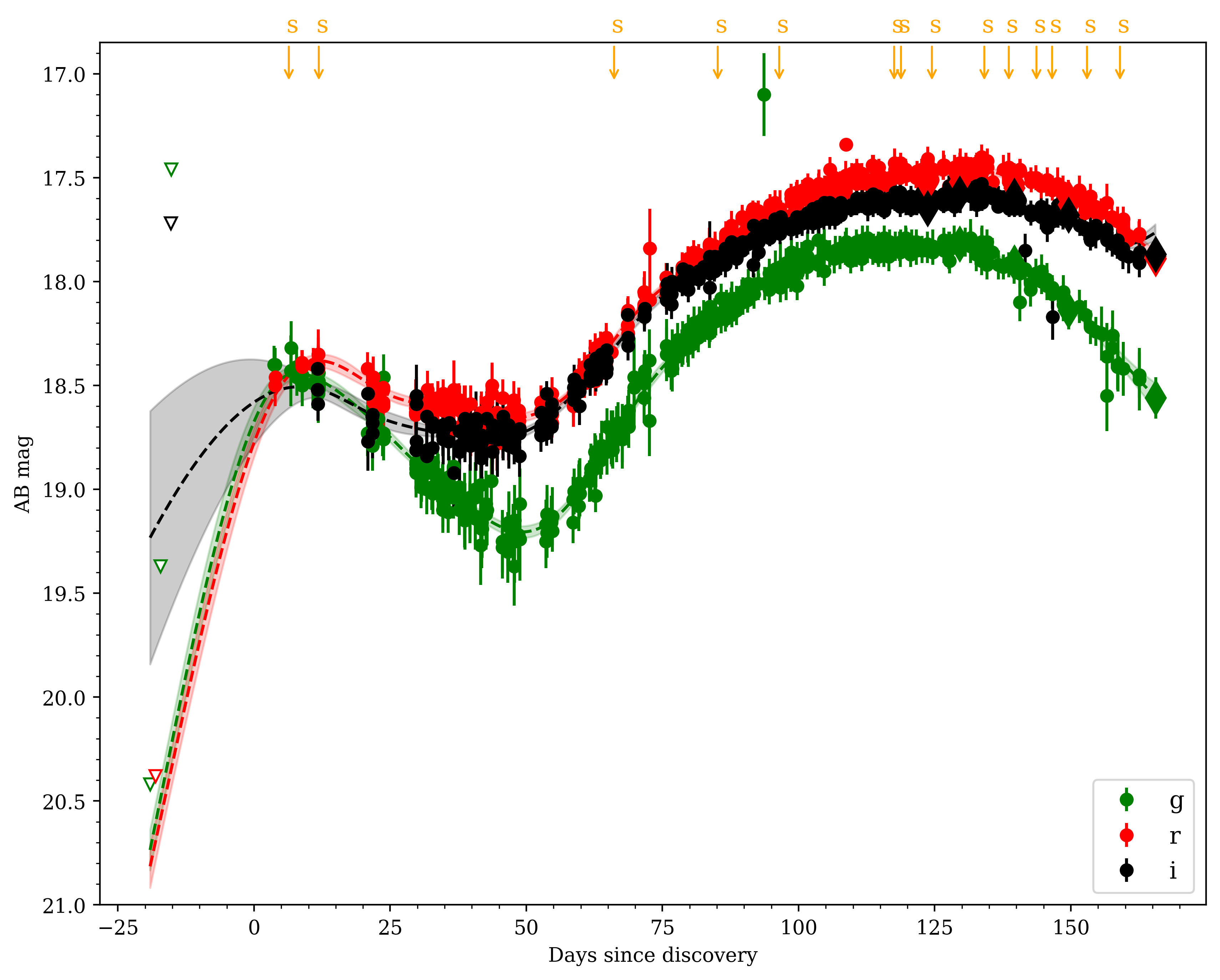}
    \caption{ 
    Light curves of \ac{obj} in $g$ (green symbols), $r$ (red) and $i$ (black) band. These are observed (AB) magnitudes plotted versus observer frame time in days since discovery. 
    The yellow downward pointing arrows on top indicate the epochs of spectroscopy, and the dashed lines with error regions are Gaussian Processing estimates of the interpolated/extrapolated \ac{lc}.
    Relevant upper limits are shown to constrain the early phase of the \ac{lc}, displayed as inverted triangles. }
    \label{fig:lc}
\end{figure*}

The reddening corrections are applied using the \cite{1989ApJ...345..245C} extinction law with $R_V=3.1$. No further host galaxy extinction has been applied, since there is no sign of any \ion{Na}{i d} absorption in our spectra. 
The \ac{lc}s are shown in Fig.~\ref{fig:lc}.

After the initial decline of about 50 days
(this is past discovery in the observer's frame), \ac{obj} started
to slowly brighten again. This continued for about 70 days and happened
in all three bands.
Once this was realized in late May 2020, a more intense follow-up
could be activated, in particular with regular spectroscopic
observations (Sect.~\ref{sec:opticalspectra}.)

We used a \ac{gp} algorithm\footnote{\href{https://george.readthedocs.io}{https://george.readthedocs.io}}
to interpolate the photometric measurements and found that the peak happened at 
m$^{\rm{peak}}_{r} = 17.49 \pm 0.01$ after 
t$^{\rm{rise}}_{r} = 114.51 \pm 0.10$ rest frame days, via $scipy.find\_peaks$. 
In the $g$ and $i$ bands the photometric behavior follows the same trend, and peaked at
m$^{\rm{peak}}_{g} = 17.83$ after 
t$^{\rm{rise}}_{g} = 114.70$ days as well as
m$^{\rm{peak}}_{i} = 17.58$ after 
t$^{\rm{rise}}_{i} = 119.70$ rest frame days.

\subsection{Swift-observations}
\label{sec:space}

\subsubsection{UVOT photometry\label{sec:uvot}}

A series of ultraviolet (UV) and optical photometry observations were obtained with the UV Optical Telescope onboard the 
{\it Swift} observatory (UVOT; \citealp{gehrels2004}; \citealp{2005SSRv..120...95R}).

Our first {\it Swift}/UVOT observation was performed on 2020 July 03 ($\mathrm{JD}=2459034.4226$) and provided detections in all the bands. However, upon inspection it is difficult to assess to what extent the emission is actually from the \ac{sn} itself, or if it is diffuse emission from the surroundings. We need to await template subtracted images to get reliable photometry.

\subsubsection{X-rays}
\label{sec:xrays}

With \swift\ we also used the onboard X-Ray Telescope (XRT;
\citealt{Burrows2005}). We analysed all data with the online-tools of the UK \swift\ team\footnote{\href{https://www.swift.ac.uk/user_objects}{https://www.swift.ac.uk/user\_objects}} that use the methods described 
in \citet{Evans2007a,Evans2009a} 
and the software package HEASoft\footnote{\href{https://heasarc.gsfc.nasa.gov/docs/software/heasoft}{https://heasarc.gsfc.nasa.gov/docs/software/heasoft}} version 6.26.1  to search for X-ray emission at the location of \ac{obj}.

Combining the five epochs taken in July 2020 amounts to a total XRT
exposure time of $\sim11000$~s ($3$~hr), and provides a $3\sigma$ upper limit of 0.001~count~s$^{-1}$ between 0.3 and 10 keV. If we assume a power-law spectrum with a photon index of $\Gamma = 2$ and a Galactic hydrogen column density of $2.65\times10^{20}$~cm$^{-2}$ \citep{HI4PI2016a} this would correspond  to an unabsorbed 0.3--10.0 keV flux of $4\times10^{-14}~{\rm erg\,cm}^{-2}\,{\rm s}^{-1}$. At the luminosity distance of \ac{obj} this corresponds to a luminosity of less than L$_{\rm{X}}=2\times10^{41}$~erg\,s$^{-1}$ (0.3--10~keV) at an epoch of $\sim107\pm9$~days rest-frame days since discovery.


\subsection{Pre-discovery imaging}
\label{sec:prediscovery}

\begin{figure*}
\centering
    \includegraphics[width=0.8\textwidth]{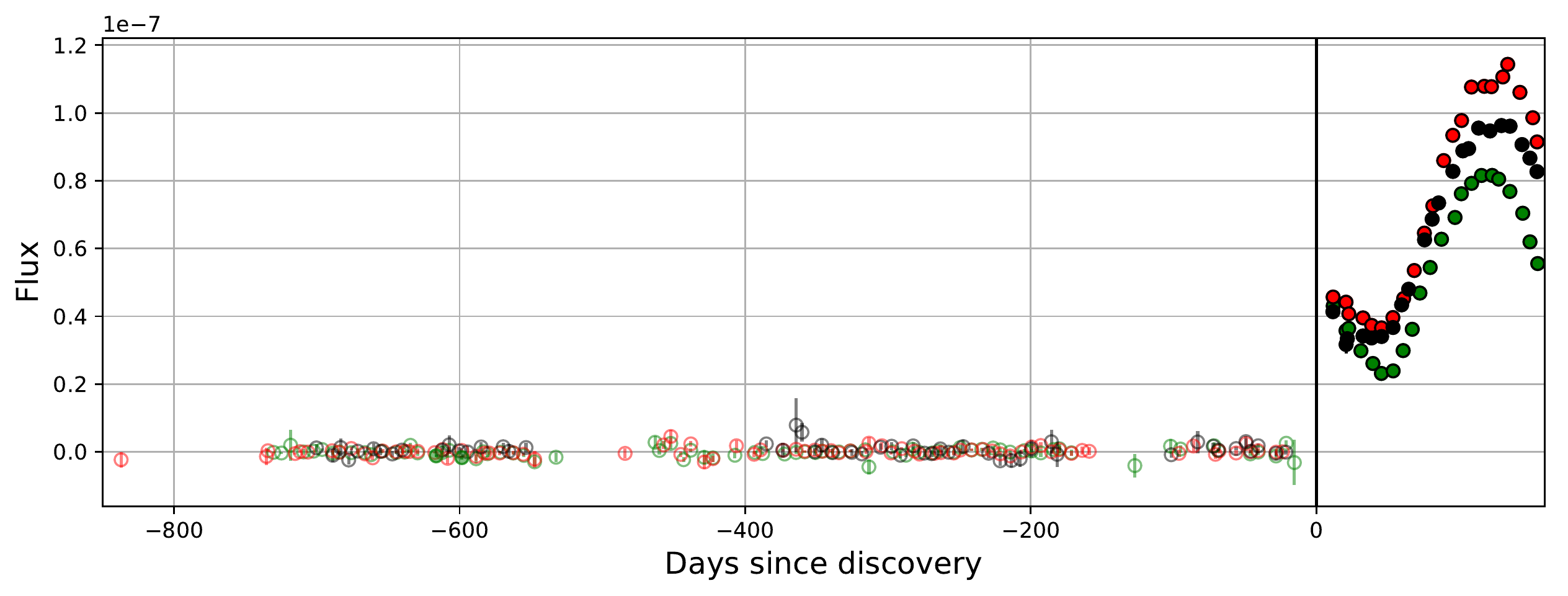}
    \caption{Pre-explosion images in \ac{ztf} for \ac{obj} reveal no precursos in $g$ (green symbols), $r$ (red) or $i$ (black) bands. 
    The flux $f$ is given as a dimensionless ratio and can be converted 
    via m$_{\rm AB}$ = $-$2.5 log$_{10}$($f$). 
    Filled data points are $\gtrsim 5 \sigma$ detections, whereas shaded points are between 3 and 5 sigma 
    and open symbols are less significant than 3 sigma.
    }
    \label{fig:precursor}
\end{figure*}

A particular peculiarity for iPTF14hls was the tentative detection of
a precursor in images taken long before the discovery of the
transient, from the year 1954. We therefore looked at the P48 imaging
of the field of \ac{obj} for some epochs prior to discovery, both by \ac{ztf}
and by the predecessor PTF.
For the PTF images, image subtraction revealed no detection ($5\sigma$) for the 65 $r$-band images obtained between May 9, 2009 and July 24, 2010. 
For \ac{ztf}, we searched for pre-explosion outbursts in 1538 observations that were obtained in the $g$, $r$ and $i$ band in the 2.3 years before the first detection of \ac{obj}. No outbursts are detected when searching unbinned or binned (1 to 90-day long bins) \ac{lc}s following the methods described by Strotjohann et al. (in prep.), see Fig.~\ref{fig:precursor}. 
The precursor detected prior to iPTF14hls had an
apparent $r$-band magnitude of $20.7$
(absolute $r$-band magnitude of $-15.6$) and we can rule out similar outbursts for 
50\% 
of the time assuming that an outburst lasts for at least one week.

\subsection{Optical spectroscopy}
\label{sec:opticalspectra}

Spectroscopic follow-up was conducted with SEDM 
mounted on the P60.
Further spectra were obtained with the Nordic Optical
Telescope (NOT) using the Alhambra Faint Object Spectrograph (ALFOSC).
A log of the spectral observations is provided in
Table~\ref{tab:spec}, which includes 14 epochs of spectroscopy.
SEDM spectra were reduced using the pipeline described by
\citet{RigaultPySEDM} and the spectra from La Palma were reduced using standard pipelines.
The spectra were finally absolute calibrated against the $r$-band magnitudes using the \ac{gp} interpolated
measured magnitudes and then corrected for MW extinction.
All spectral data and corresponding information will be made available via WISeREP\footnote{\href{https://wiserep.weizmann.ac.il}{https://wiserep.weizmann.ac.il}} \citep{wiserep}.

\subsection{Host galaxy}
\label{sec:host}
\subsubsection{Photometry}

We retrieved science-ready coadded images from the \textit{Galaxy Evolution Explorer} (\galex) general release 6/7 \citep{Martin2005a}, the Panoramic Survey Telescope and Rapid Response System (Pan-STARRS, PS1) Data Release 1 \citep{Chambers2016a}, the Two Micron All Sky Survey \citep[2MASS;][]{Skrutskie2006a}, and preprocessed \wise\ images \citep{Wright2010a} from the unWISE archive \citep{Lang2014a}\footnote{\href{http://unwise.me}{http://unwise.me}}. The unWISE images are based on the public \wise\ data and include images from the ongoing NEOWISE-Reactivation mission R3 \citep{Mainzer2014a, Meisner2017a}.

We measured the brightness of the host in a consistent way from the far-ultraviolet to the mid-infrared (i.e., measuring the total flux and preserving the instrinsic galaxy colours) using LAMBDAR\footnote{\href{https://github.com/AngusWright/LAMBDAR}{https://github.com/AngusWright/LAMBDAR}} \citep[Lambda Adaptive Multi-Band Deblending Algorithm in R;][]{Wright2016a} and the methods described in \citet{Schulze2020a}. Table~\ref{tab:host} gives the measurements in the different bands.

\subsubsection{Spectral energy distribution modelling}
\label{sec:hostsedm}

We modelled the spectral energy distribution with the software package prospector version 0.3 \citep{Leja2017a}. Prospector uses the Flexible Stellar Population Synthesis (FSPS) code \citep{Conroy2009a} to generate the underlying physical model and python-fsps \citep{ForemanMackey2014a} to interface with FSPS in python. The FSPS code also accounts for the contribution from the diffuse gas (e.g., HII regions) based on the Cloudy models from \citet{Byler2017a}. 
We assumed a Chabrier initial mass function \citep{Chabrier2003a} and approximated the star formation history (SFH) by a linearly increasing SFH at early times followed by an exponential decline at late times.
The model was attenuated with the \citet{Calzetti2000a} model. Finally, we use the dynamic nested sampling package dynesty \citep{Speagle2020a} to sample the posterior probability function.

\subsubsection{Host galaxy spectrum}
\label{hostgalaxyspectrum}

To obtain a spectrum of the host galaxy, we aligned the slit along the host galaxy in our latest NOT observation. In this way we could also extract a
spectrum 
from the region around the \ac{sn} position 
in order to measure the host galaxy emission lines. This was done to carry out an abundance analysis based on the strong line methods. 

\section{Analysis and Discussion}
\label{sec:discussion}


\subsection{Light curves}
\label{sec:lc}

The $g$-, $r$- and $i$-band \ac{lc}s of our \ac{sn} are displayed in
Fig.~\ref{fig:lc}.
The general behaviour of the \ac{lc}s was already discussed in
Sect.~\ref{sec:optical}, and the main characteristic is of course the slow evolution with the
initial decline followed by the late rise over several months. In the figure we have also
included the most restrictive upper limits as triangles ($5\sigma$),
while the arrows on top of the figure illustrate epochs of
spectroscopy. The \ac{gp} interpolation is also shown, which is used
for absolute calibrating the spectra.
For the \ac{gp}, we perform time series forecasting for the joint multi-band fluxes with their corresponding central wavelengths, in order to include colour information. Here, we use a flat mean function and a stationary kernel {\tt Matern 3/2} for the flux form. 
\ac{gp} is also used elsewhere in this work to interpolate data, and we then 
use the flux model from \cite{villar} for the mean function.

\begin{figure*}
\centering
    \includegraphics[width=0.8\textwidth]{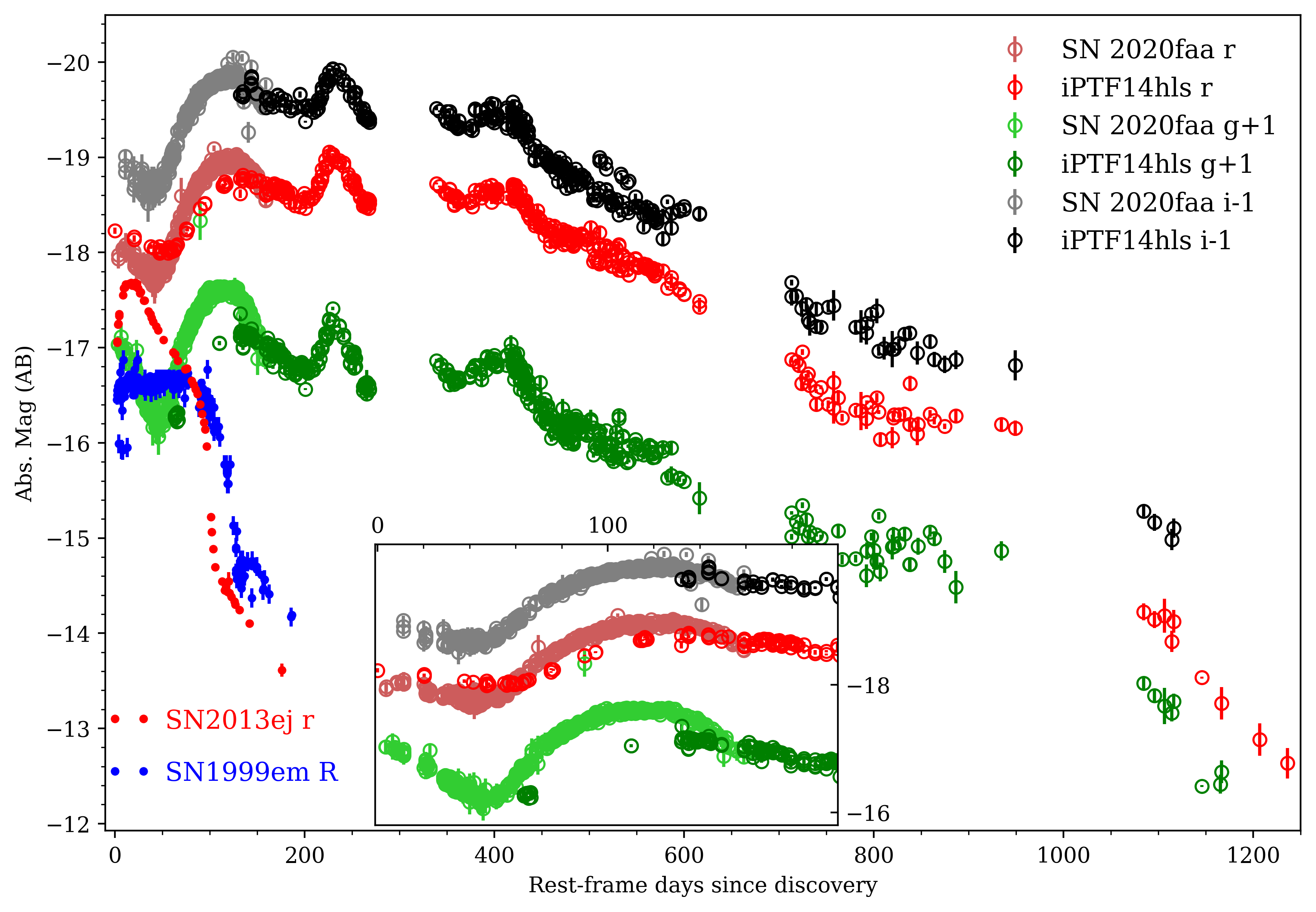}
    \caption{Absolute magnitudes of \ac{obj} together with the \ac{lc}s of iPTF14hls, as well as SNe 1999em and 2013ej. No scaling has been applied to match these \ac{sne}. The inset highlights the early evolution (exactly 200 days), 
    which is where \ac{obj} demonstrates a striking similarity with the early iPTF14hls \ac{lc}s.}
    \label{fig:lc+iptf}
\end{figure*}

In Fig.~\ref{fig:lc+iptf} we show the $g$-, $r$- and $i$-band \ac{lc}s in
absolute magnitudes together with the \ac{lc}s of iPTF14hls
from S19. The bottom part has an inset highlighting the first 200 days, which 
zooms in on the evolution of \ac{obj}.
The magnitudes in Fig.~\ref{fig:lc+iptf} are in the AB system and have been
corrected for distance modulus and MW extinction, and are plotted
versus rest frame days past discovery.

The inset shows the remarkable similarity in absolute magnitude and
timescale of the two \ac{sne}, whereas the full figure might be seen as a
prediction for the future evolution of \ac{obj}. We will continue to
follow the \ac{sn} at best effort with \ac{ztf}, but report on these results
already now to encourage also the community to keep an eye on the continued
evolution of this transient. We note that with a declination of $+72$
degrees the source is well placed to be observed around the year from
Northern observatories. 
No offset was applied to match the absolute magnitudes, they fall very well on top of each other anyway. Note that also no shift was applied in the time scales, we have plotted iPTF14hls since time of discovery, which supports a similar evolution also in this dimension. It is worth to note that the explosion date\footnote{or maybe better, time of first light.} for iPTF14hls was unconstrained by several months (A17), which made it more difficult to estimate for example total radiated energy for that \ac{sn}. The comparison here makes it likely that iPTF14hls was not discovered very late after all.

Needless to say, the evolution is very different from that of normal \ac{sne} Type II, which was already demonstrated by A17 in the comparison of iPTF14hls to SN 1999em 
(see also Fig.~\ref{fig:lc+iptf}).
A Type II \ac{sn} normally stays on a relatively flat plateau for about 100 days, and then quickly plummets to the radioactive decay tail. The rejuvenated long-timescale rise for \ac{obj} argues, as for iPTF14hls, that a different powering mechanism must be at play. 

\begin{figure*}
\centering
    \includegraphics[width=0.8\textwidth]{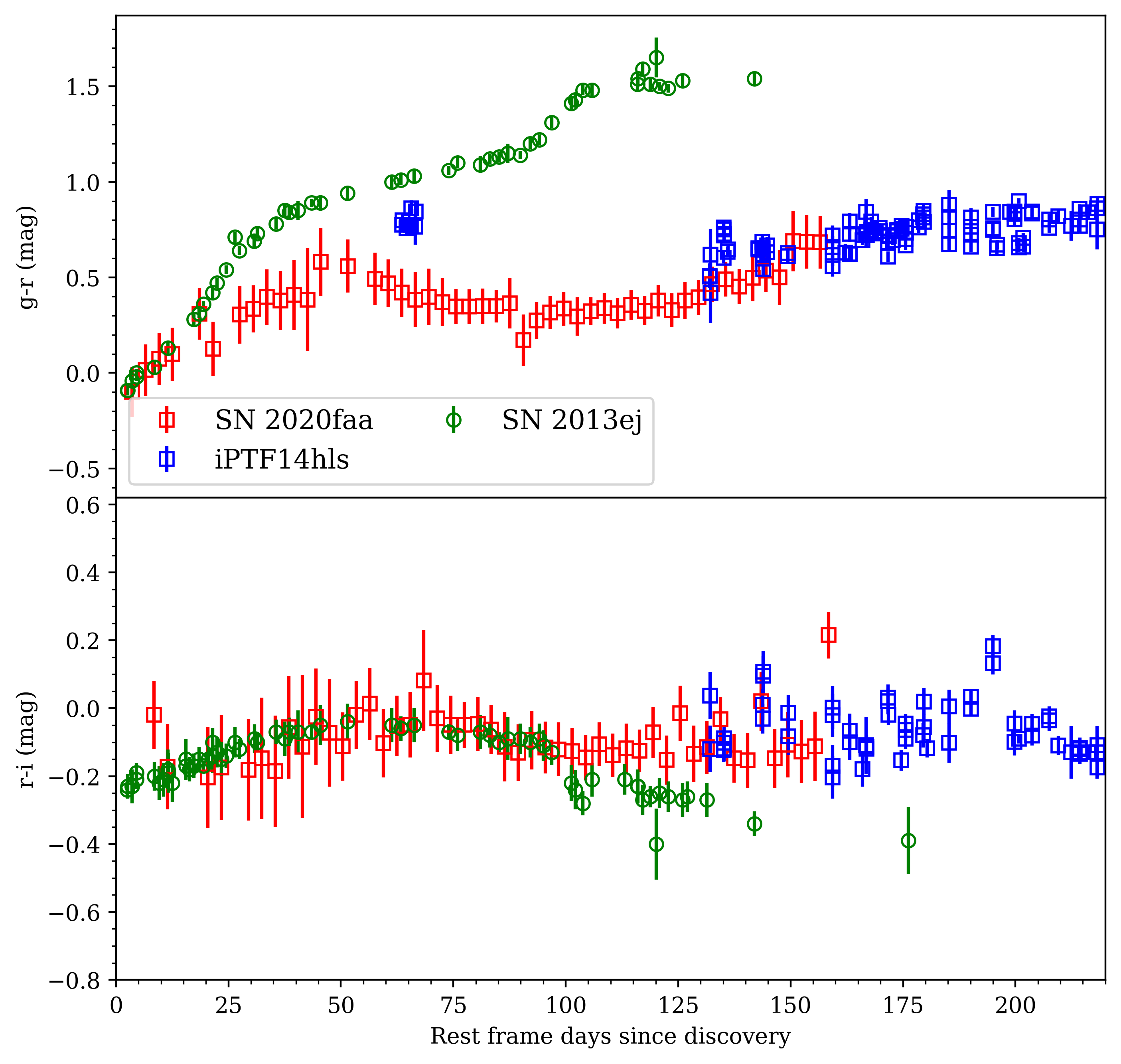}
    \caption{Colour evolution of \ac{obj} shown in $g-r$ (upper panel) and $r-i$ (lower panel), binned in 3 days. The colours have been corrected for MW extinction and are plotted in rest frame days relative to epoch of discovery. For comparison we have also plotted colours for iPTF14hls and for one normal Type II \ac{sn} 2013ej, whose epochs are also provided in rest frame days since discovery.}
    \label{fig:color}
\end{figure*}

The colour evolution of \ac{obj} is shown in Fig.~\ref{fig:color}.
We plot $g-r$ in the upper panel and $r-i$ in the lower panel, both corrected for MW extinction. In doing this, no interpolation was used. Given the excellent \ac{lc} sampling we used only data where the pass-band magnitudes were closer in time than 0.1 days.
Comparison is made with the colour evolution for iPTF14hls, but this \ac{sn} was not covered at early phases. There is anyway evidence for similar colours, which argue against significant host extinction. We also compare the colours against a normal Type II \ac{sn} 2013ej \citep{Valenti_2013}, which is selected because it is also not suffering from host-galaxy reddening and has been well monitored in the $gri$ photometric system.
These photometric data were obtained via the 
\href{https://sne.space/}{Open Supernova Catalog}. 
As shown, \ac{obj} is relatively bluer than  the normal Type II SN in $g-r$ colour space, whereas they are matching well in $r-i$.

\subsection{Spectra}
\label{sec:spectra}

\begin{figure*}
\centering
    \includegraphics[width=\textwidth]{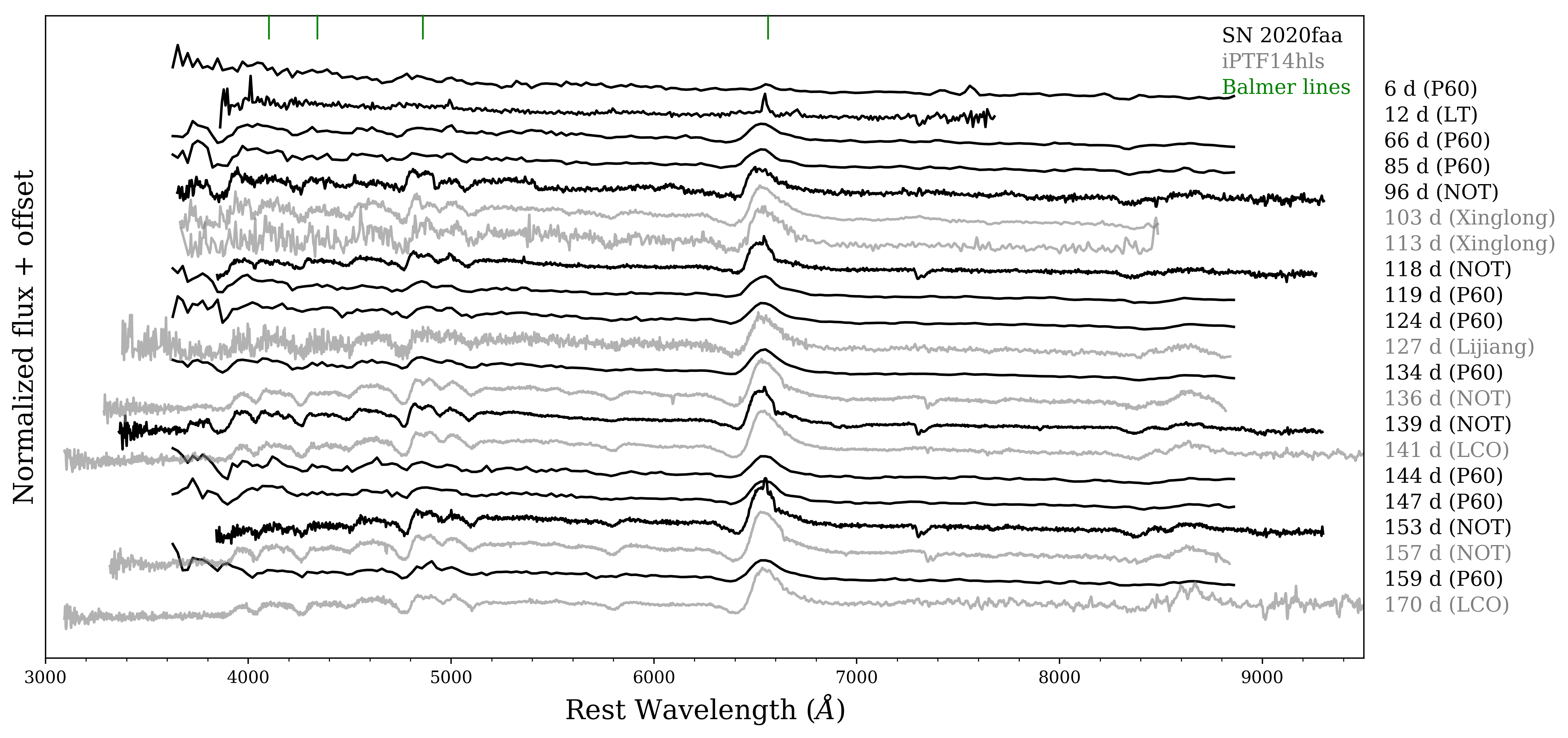}
    \caption{Sequence of optical spectra for \ac{obj}. The complete log of spectra is provided in Table~\ref{tab:spec}. The epoch of the spectrum is provided to the right. For comparison we also show spectra of iPTF14hls in grey.}
    \label{fig:spectra}
\end{figure*}

The log of spectroscopic observations was provided in 
Table~\ref{tab:spec}
and the sequence of spectra is shown in Fig.~\ref{fig:spectra}.
Overall, these are spectra of a typical Type II \ac{sn}. We
compare these with spectra from iPTF14hls. Note that the rise of
iPTF14hls was not picked up immediately and therefore the first
spectrum of that \ac{sn} was only obtained more than 100 days past first detection. 
We were faster for \ac{obj}, i.e. the first P60 spectrum was obtained 6 days past discovery, and we can measure the evolution of the expansion
velocity from 65 days past discovery.  

\begin{figure*}
\centering
    \includegraphics[width=0.8\textwidth]{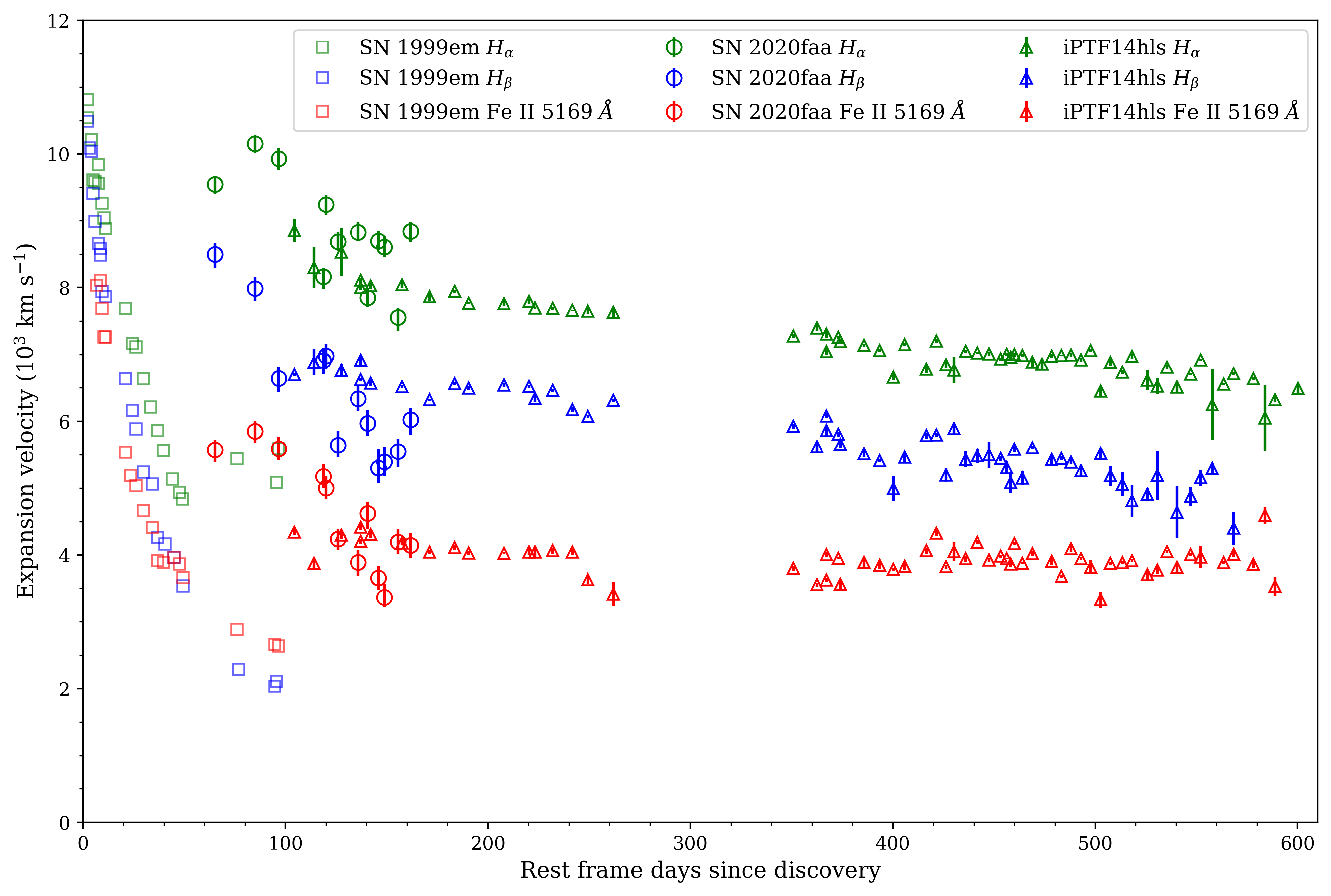}
    \caption{Velocities estimated from the P-Cygni minima of H$\alpha$ (green), H$\beta$ (blue) and \ion{Fe}{ii} $\lambda$5169 (red) for the three \ac{sne} discussed throughout the paper. Whereas the normal Type II \ac{sn} 1999em shows a fast decline in expansion velocities, iPTF14hls exhibits virtually constant velocities, where the Fe velocity was significantly lower than those estimated from Balmer lines at all epochs. For \ac{obj} we probe intermediate phases and see a slowing down of the photosphere, but with velocities very similar to those demonstrated by iPTF14hls at the common epochs around 150 days.}
    \label{fig:velocity}
\end{figure*}

These velocities are shown in Fig.~\ref{fig:velocity}, where
we compare to iPTF14hls and to SN 1999em following the methodology of
A17, see their fig.~3. We measured the velocities for \ac{obj} using {\tt iraf} to fit a Gaussian to the minimum of the absorption lines of the corresponding P-Cygni profiles.
The difference in rest wavelength between the minimum of the best-fit Gaussian to the line location was translated to an expansion velocity.
Uncertainties in the velocities were estimated by a random sampling on the Gaussian minimum 1000 times by $\pm 5 \AA$, as in A17.
The time evolution of the velocities measured for H$\alpha$, H$\beta$ and for 
\ion{Fe}{ii} $\lambda$5169 match well with those of iPTF14hls at the common epochs, but also extend to earlier phases. The velocities for the comparison are taken from A17. The striking characteristic of the time evolution for iPTF14hls was the very flat velocity evolution. We do not know (yet) if \ac{obj} will follow such a flat evolution, or if iPTF14hls had a faster evolution in the first 100 days.

The velocities for the first two \ac{obj} spectra are not presented in Fig.~\ref{fig:velocity}. The first P60 does not have enough signal, and there is no
P-Cygni profile in the LT spectrum.
However, according to the widths of the emission components of H$\alpha$ and H$\beta$, we estimate the expansion velocities 
are $\sim$9000 km s$^{-1}$, consistent with those inferred later from the P-Cygni minimum.

\begin{figure*}
\centering
    \includegraphics[width=0.8\textwidth]{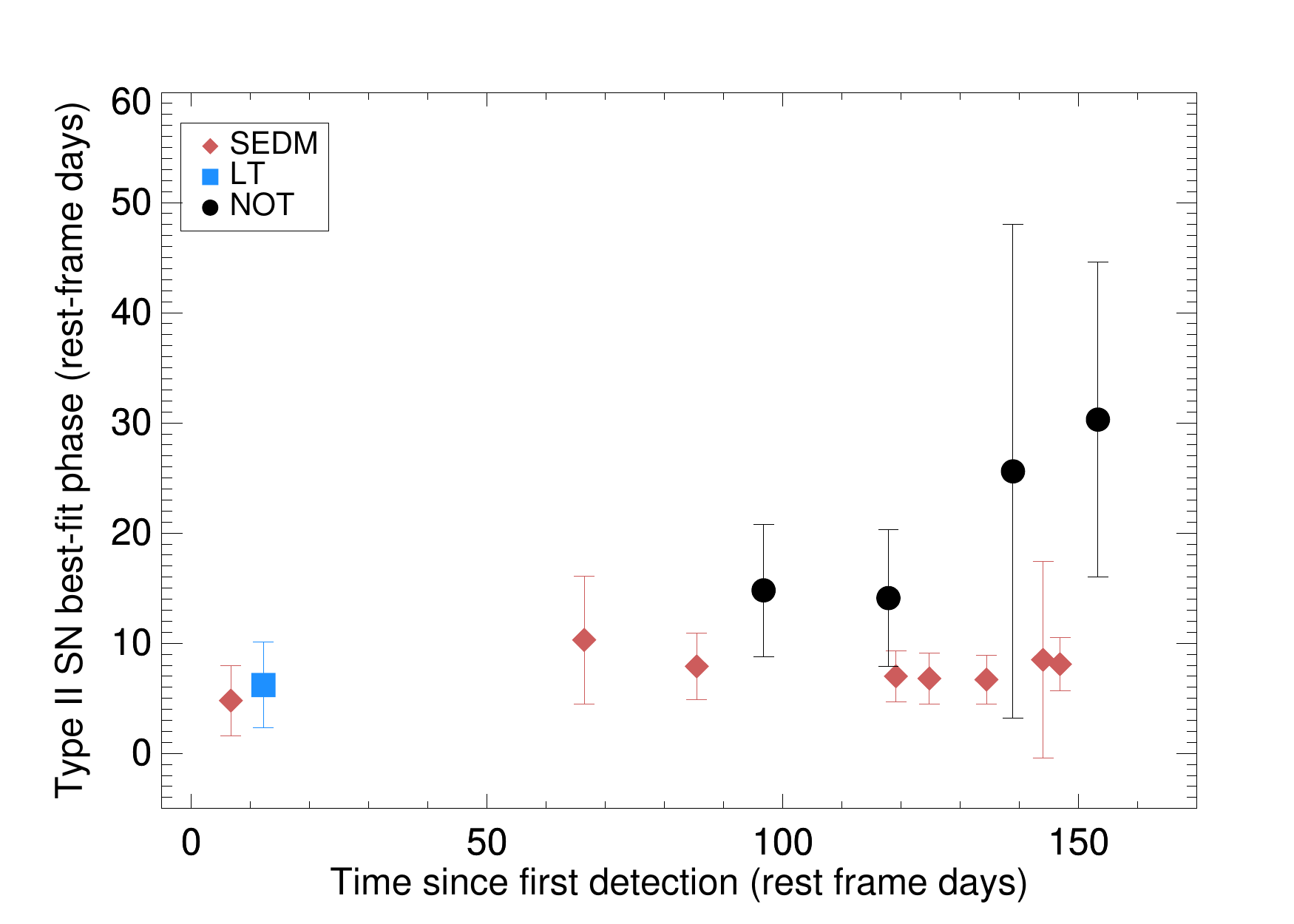}
    \caption{Phases estimated by comparison to superfit templates are plotted versus rest frame days since first detection for \ac{obj}. The overall spectral evolution revealed by these comparisons is very slow and even at more than 100 days the best matches are with younger Type II \ac{sne}. This is similar to what was found by A17 for iPTF14hls, which continued to display slow evolution for 600+ days.}
    \label{fig:superfit_phase}
\end{figure*}

A striking characteristic of iPTF14hls was its very slow spectral evolution, with the spectral phase inferred from comparing to typical \ac{sne}~II such as SN 1999em being a factor of $6-10$ younger than the true 
phase. To check whether the same is true for \ac{obj}, we follow A17 and use the spectral comparison code {\tt superfit} \citep{Howell2005} to estimate the phase of each spectrum against a library of \ac{sn}~II templates. The results are shown in Fig.~\ref{fig:superfit_phase}, with the estimated spectral age plotted versus actual age. The evolution is remarkably flat, and in particular all the spectra taken during the rebrightening phase to date ($50-150$ days past discovery) match \ac{sn} templates with phases
$7-30$ days past peak light. This is similar to what was seen in A17 (compare their Extended Data fig.~4), and thus, slow spectral evolution is another property that iPTF14hls and \ac{obj} have in common.


\subsection{Bolometric lightcurve}
\label{sec:bb}

In order to estimate a total luminosity radiative output, we attempted to construct a bolometric \ac{lc}.
We follow a similar Black-body (BB) approximation approach as done for iPTF14hls by A17, 
and for the early evolution probed here we have better photometric colour coverage to pursue this.

\begin{figure*}
\centering
    \includegraphics[width=0.8\textwidth]{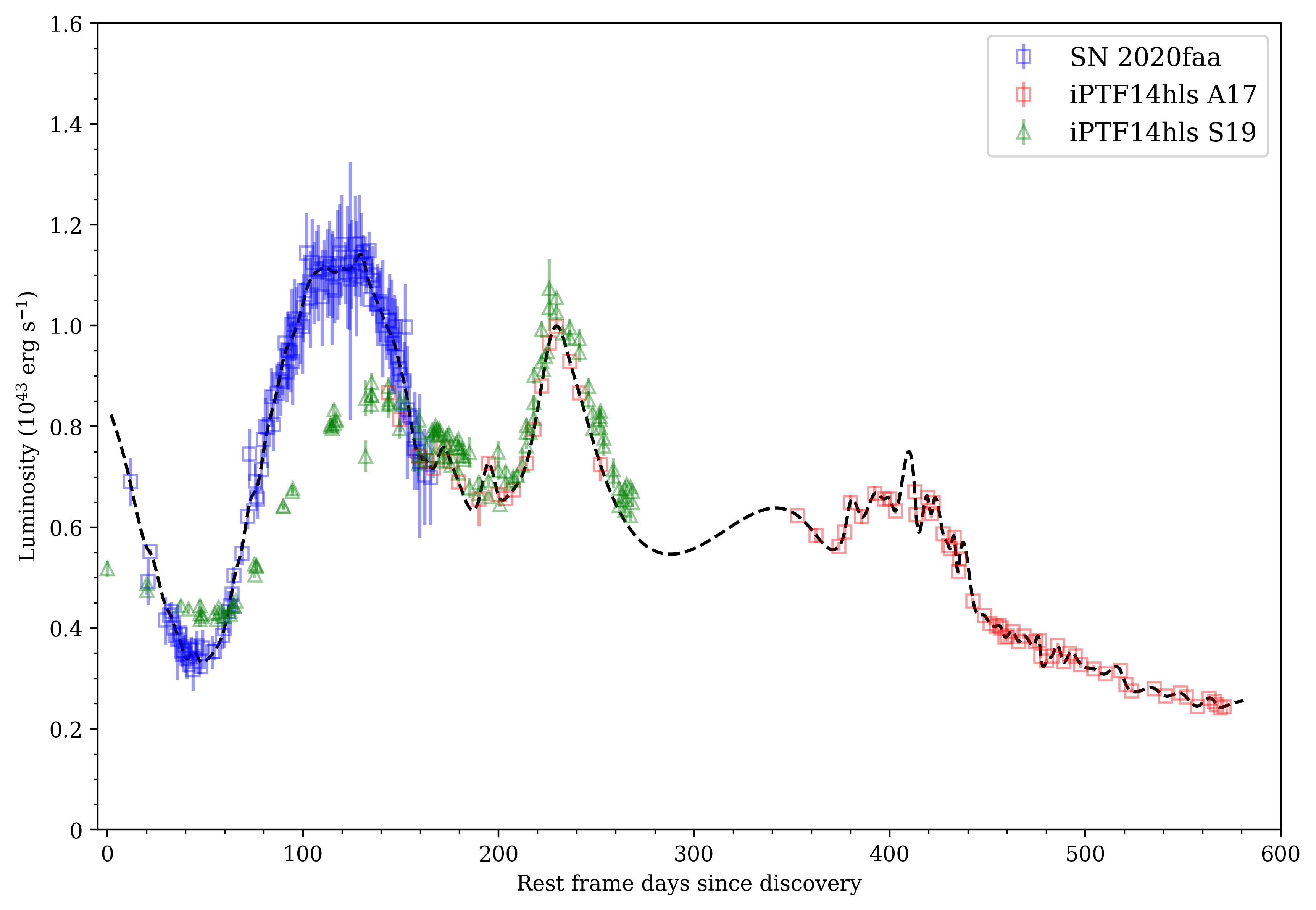}
    \caption{Luminosity of \ac{obj} after accounting for MW extinction, distance and integrating a BB fit to the $gri$ photometry. A similar method was used for iPTF14hls which only had colour data past 150 days, and we can see that the early time emission of \ac{obj} nicely merges with the late time luminosity for iPTF14hls.
    The \ac{gp} fit on the joint \ac{lc}s of \ac{obj} and iPTF14hls is shown as a black dashed line. 
    In green we show the luminosity estimate for iPTF14hls from S19 up to 300 days, which assumed a constant bolometric correction at early times.}
        \label{fig:luminosity}
\end{figure*}

The result is shown in Fig.~\ref{fig:luminosity}. 
The red squares show the luminosity of 
iPTF14hls (from A17, their Extended Data fig. 2). There was only enough colour information to fully construct this luminosity for iPTF14hls at later epochs. For \ac{obj}, we can use the $gri$ coverage to estimate the luminosity before this as well, and see that those estimates connect nicely at 150 days post discovery. 
Using this, we can estimate a
maximum bolometric luminosity for \ac{obj} of
L$_{\rm{bol}}$= $1.12 \times 10^{43}$ erg s$^{-1}$
at 120.6 rest frame days
and a total radiated energy over the first 162.4 rest frame days (until the end of observations presented in this paper, i.e. 2020 September 06) of
E$_{\rm{rad}}$= $1.0 \times 10^{50}$ erg. 
This can be compared to the total radiative output of iPTF14hls which was
E$_{\rm{rad}}$= $3.59 \times 10^{50}$ erg over 1235 days (S19). In that paper, the early bolometric of iPTF14hls was reconstructed, and that comparison is also shown in Fig.~\ref{fig:luminosity}. Within the uncertainties, these are quite similar, the S19 early bolometric luminosity was estimated from the $r$-band data and a constant bolometric correction. 
Already the first 160 days of \ac{obj} cannot easily be powered by the mechanism usually responsible for a Type II \ac{sn} \ac{lc} - radioactive decay. Using, 
$L = 1.45 \times 10^{43}$ exp$(-\frac{t}{\tau_{Co}}) (\frac{M_{Ni}}{M_{\sun}}) $ erg s$^{-1}$ from \cite{Nadyozhin} implies that we would require 
a solar mass
of $^{56}$Ni to account for the energy budget. This is already out of the scope for the traditionally considered neutrino explosion mechanism \cite[e.g.,][]{Terreran}.


\begin{figure*}
\centering
    \includegraphics[width=0.8\textwidth]{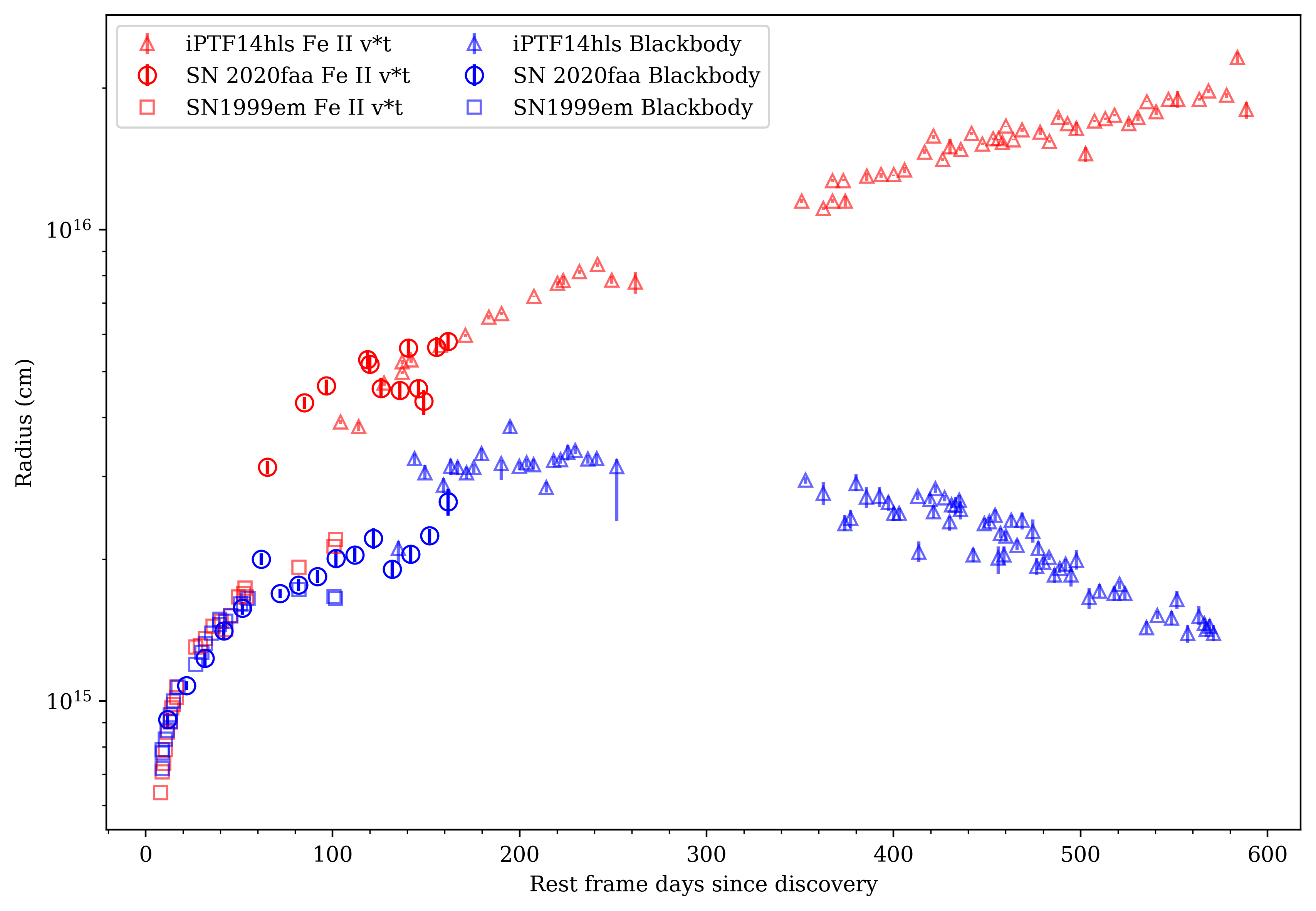}
    \caption{Evolution of the the radius as a function of time for \ac{obj} (binned in 10 days), as compared to the extraordinary iPTF14hls and the regular Type II \ac{sn} 1999em. This figure closely follows the presentation from A17, their fig. 4, and shows estimates for the radius evolution from two different methods for the three different \ac{sne}. 
    A main theme in A17 was that for iPTF14hls, the radius evolution estimated from the BB approximation and the radius estimated from the spectroscopic velocities were different and diverged with time. }
    \label{fig:radius}
\end{figure*}

From the BB approximation we also obtain the temperature and the evolution of the BB radius. The radius evolution was an important clue to the nature of iPTF14hls in A17 (their fig. 4), and we therefore show a very similar plot in 
Fig.~\ref{fig:radius}.  
The radius thus obtained is directly compared to the values for iPTF14hls and 
SN~1999em. We here also include the radius estimated from the spectroscopic velocities, estimated from the P-Cygni minima of the \ion{Fe}{II} $\lambda$5169 line.

The figure shows that the BB radius of \ac{obj} at the earliest phases are similar and evolve similarly to those of SN 1999em, and approach the values of the radius for iPTF14hls at 140 days. The $vt$ radius on the other hand are higher for \ac{obj}, just as they were for iPTF14hls. We can see that they smoothly attach to the values for iPTF14hls.

\subsection{Host galaxy}

\begin{figure}
\centering
    \includegraphics[width=1\columnwidth]{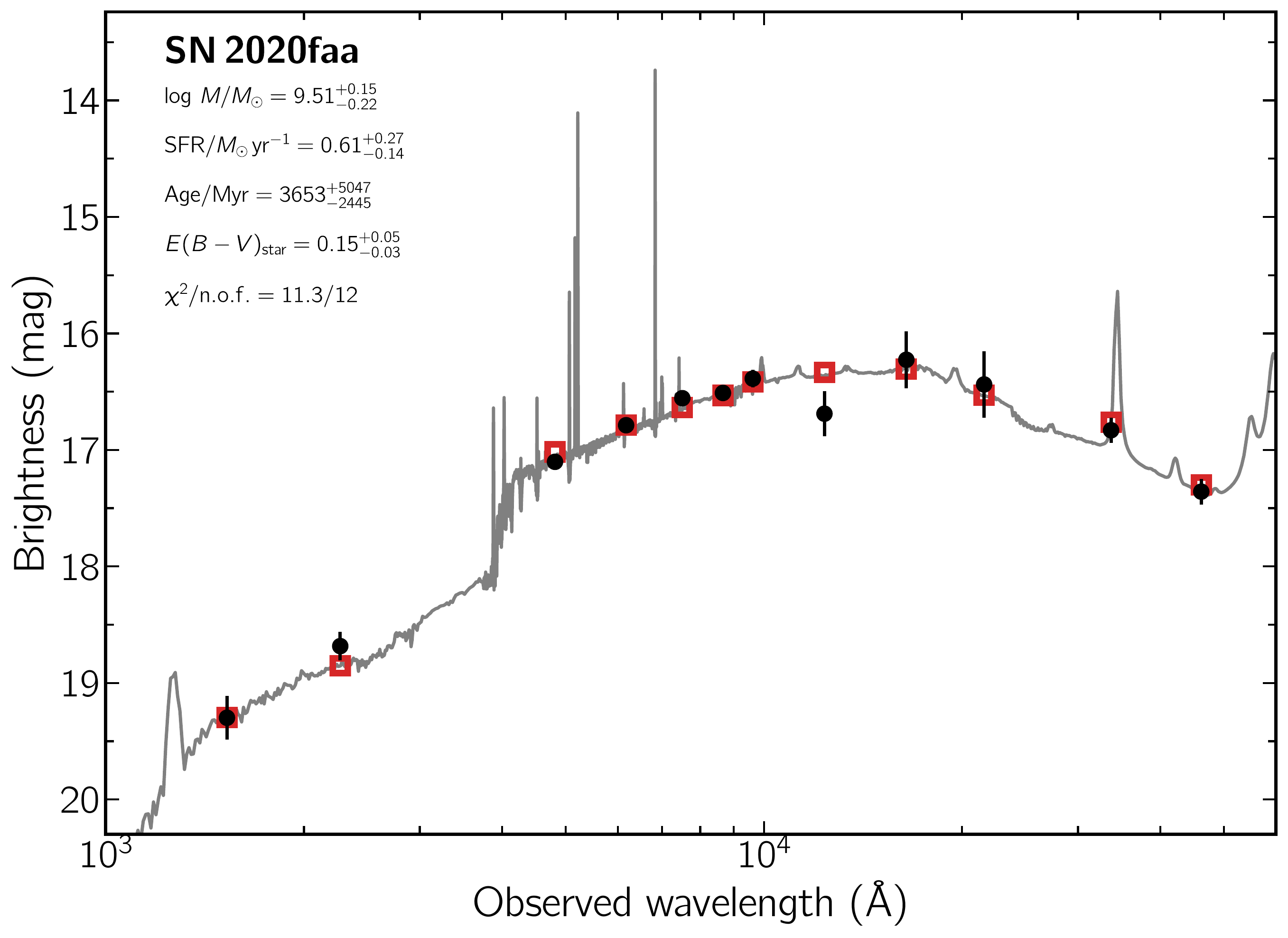}
    \caption{Spectral energy distribution (SED) of the host galaxy of \ac{obj} from 1000 to 60,000 \AA\ (black data points). The solid line displays the best-fitting model of the SED. The red squares represent the model-predicted magnitudes. The fitting parameters are shown in the upper-left corner. The abbreviation ``n.o.f.'' stands for numbers of filters.}
\label{fig:host_sed}
\end{figure}

\begin{figure}
\centering
    \includegraphics[width=1\columnwidth]{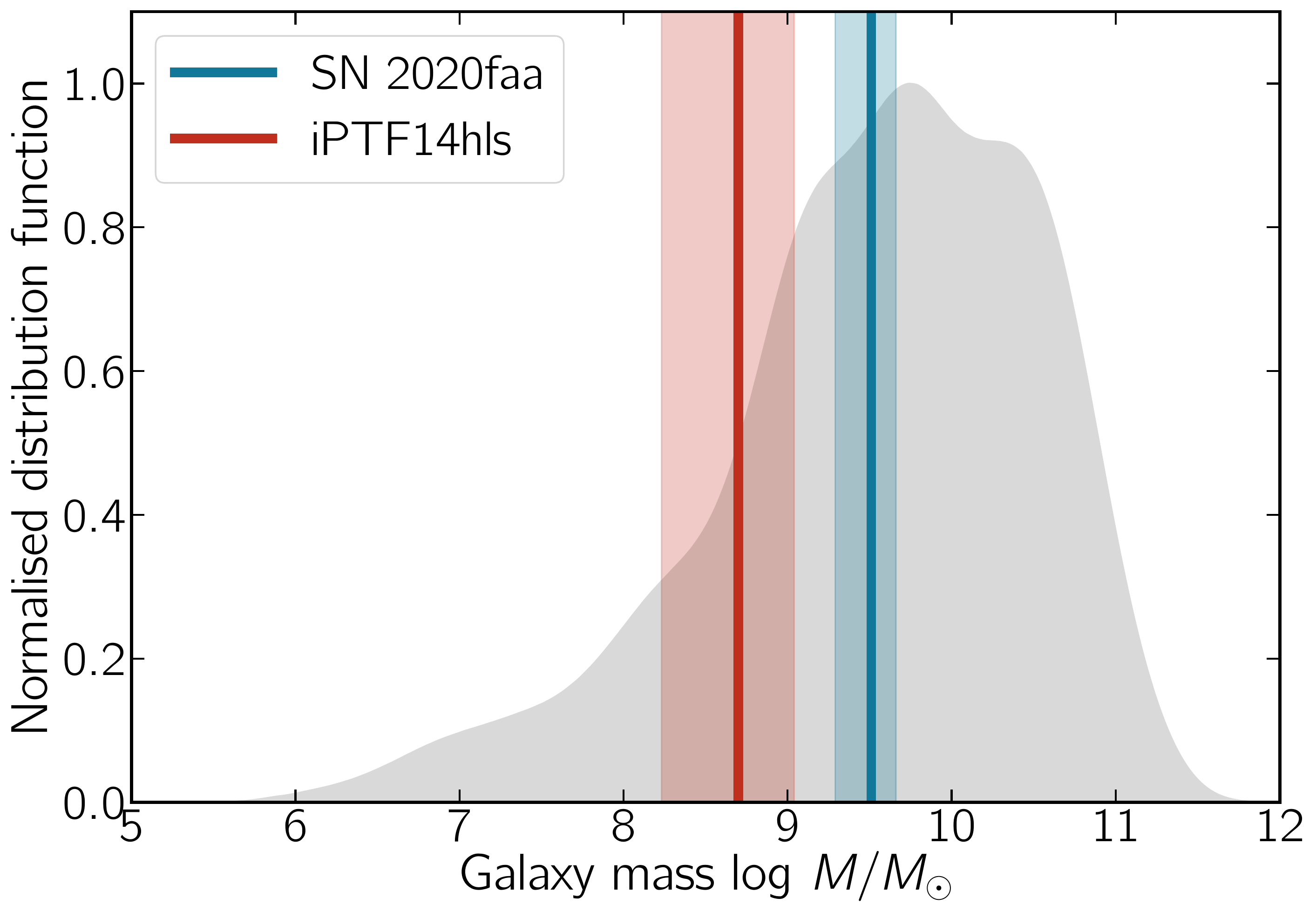}
    \caption{The host-galaxy mass of \ac{obj} and iPTF14hls in the context of host galaxies for \ac{sne} II from the PTF and iPTF survey (as presented by \citealt{Schulze2020a}).}
\label{fig:host_sed2}
\end{figure}

The results of the SED modeling (Sect. \ref{sec:hostsedm}) of the host galaxy is displayed in Fig.~\ref{fig:host_sed}. We obtain a good fit for a galaxy with a mass of 3.2$\times$10$^{9}$ \msun $~$ and a star-formation rate of 0.6 \msun $~$ per year. This is a relatively regular host galaxy for a Type II \ac{sn}. In Fig.~\ref{fig:host_sed2} we compare the host mass with the distribution of host masses for \ac{sne} II from the PTF survey from \cite{Schulze2020a}. As can be seen, the host of \ac{obj} is a regular host galaxy in this respect, and is slightly more massive than the host of iPTF14hls, which is also illustrated in the figure.

\subsubsection{Host galaxy metallicity}

Using the emission lines from Sect. \ref{hostgalaxyspectrum}, we can
adopt the N2 scale of the \cite{Pettini_2004} calibration using the flux ratio between [\ion{N}{II}] $\lambda$6583 and H$\alpha$. We found that $12+{\rm log (O/H)}=8.50\pm0.15$.
We also employed another metallicity diagnostic of 
\cite{Dopita_2016} using [\ion{N}{II}], H$\alpha$ and [\ion{S}{II}] lines, in which $12+{\rm log(O/H)}=8.40\pm0.10$.   
Assuming a solar abundance of 8.69 \citep{Asplund_2009}, the oxygen abundance of the host galaxy is  $0.63^{+0.26}_{-0.12}$\,$Z_{\odot}$. Comparing to the stellar mass estimate ($10^{9.51}$\,$M_{\odot}$), our metallicity is consistent with the galaxy mass-metallicity relation (e.g. \citealt{Andrews_2013}).

\section{Summary and Conclusions}
\label{sec:summary}

We have presented \ac{obj}, a young sibling to the spectacular iPTF14hls. The first 150 days of the \ac{lc} evolution is very different from a normal Type II supernova, and very similar to that of iPTF14hls. We therefore encourage continued monitoring of this transient to explore if it will evolve in a similar fashion, with \ac{lc} undulations, 
longevity and a slow spectral evolution.  From the observations already in hand, we can conclude that just as for iPTF14hls the energy budget is already too high to be driven by a standard radioactivity scenario. The plethora of other powering mechanisms needs to be dusted off again, to explain the evolution of \ac{obj}.

\ac{ztf} will continue operations as ZTFII, with more discoveries in sight. Several community brokers are already processing the data in real time  and more activity is foreseen as we come closer to the era of the Vera Rubin telescope. The broker {\tt Alerce} \citep{alerce} is an example where a combination of computer filtering and human inspection already provides early alerts for infant \ac{sne}. We also need to keep an eye on \ac{sn} \ac{lc}s that behave in unusual and interesting ways also at later stages. This includes re-brightenings as for \ac{obj} here or due to late CSM interaction as in \cite{Sollerman2020}, 
 but could also be rapid declines or undulations, as in iPTF14hls. Hitherto most of these have been found by human scanners reacting to a 'funny' \ac{lc}. This will unlikely be the case in the Rubin era.

\begin{acknowledgements}

Based on observations obtained with the Samuel Oschin Telescope 48-inch and the 60-inch Telescope at the Palomar Observatory as part of the Zwicky Transient Facility project. ZTF is supported by the National Science Foundation under Grant No. AST-1440341 and a collaboration including Caltech, IPAC, the Weizmann Institute for Science, the Oskar Klein Center at Stockholm University, the University of Maryland, the University of Washington, Deutsches Elektronen-Synchrotron and Humboldt University, Los Alamos National Laboratories, the TANGO Consortium of Taiwan, the University of Wisconsin at Milwaukee, and Lawrence Berkeley National Laboratories. Operations are conducted by COO, IPAC, and UW.  The SED Machine is based upon work supported by the National Science Foundation under Grant No. 1106171. This work was supported by the GROWTH \citep{growth} project funded by the National Science Foundation under Grant No 1545949.

Partially based on observations made with the Nordic Optical Telescope, operated at the Observatorio del Roque de los Muchachos, La Palma, Spain, of the Instituto de Astrofisica de Canarias. Some of the data presented here were obtained with ALFOSC, which is provided by the Instituto de Astrofisica de Andalucia (IAA) under a joint agreement with the University of Copenhagen and NOTSA.
S. Y. and E. K. are funded through the GREAT research environment grant 2016-06012, and E. K. acknowledges support from The Wenner-Gren Foundations. R.L. is supported by a Marie Sk\l{}odowska-Curie Individual Fellowship within the Horizon 2020 European Union (EU) Framework Programme for Research and Innovation (H2020-MSCA-IF-2017-794467). T.-W. C. acknowledges the EU Funding under Marie Sk\l{}odowska-Curie grant H2020-MSCA-IF-2018-842471.
Thanks to David Kaplan for a careful read.
\end{acknowledgements}

\bibliography{ref}

\begin{thebibliography}{57}
\expandafter\ifx\csname natexlab\endcsname\relax\def\natexlab#1{#1}\fi

\bibitem[{{Ahn} {et~al.}(2014){Ahn}, {Alexandroff}, {Allende Prieto}, {Anders},
  {Anderson}, {Anderton}, {Andrews}, {Aubourg}, {Bailey}, {Bastien},
  {Bautista}, {Beers}, {Beifiori}, {Bender}, {Berlind}, {Beutler}, {Bhardwaj},
  {Bird}, {Bizyaev}, {Blake}, {Blanton}, {Blomqvist}, {Bochanski}, {Bolton},
  {Borde}, {Bovy}, {Shelden Bradley}, {Brand t}, {Brauer}, {Brinkmann},
  {Brownstein}, {Busca}, {Carithers}, {Carlberg}, {Carnero}, {Carr},
  {Chiappini}, {Chojnowski}, {Chuang}, {Comparat}, {Crepp}, {Cristiani},
  {Croft}, {Cuesta}, {Cunha}, {da Costa}, {Dawson}, {De Lee}, {Dean},
  {Delubac}, {Deshpande}, {Dhital}, {Ealet}, {Ebelke}, {Edmondson},
  {Eisenstein}, {Epstein}, {Escoffier}, {Esposito}, {Evans}, {Fabbian}, {Fan},
  {Favole}, {Femen{\'\i}a Castell{\'a}}, {Fern{\'a}ndez Alvar}, {Feuillet},
  {Filiz Ak}, {Finley}, {Fleming}, {Font-Ribera}, {Frinchaboy},
  {Galbraith-Frew}, {Garc{\'\i}a-Hern{\'a}ndez}, {Garc{\'\i}a P{\'e}rez}, {Ge},
  {G{\'e}nova-Santos}, {Gillespie}, {Girardi}, {Gonz{\'a}lez Hern{\'a}ndez},
  {Gott}, {Gunn}, {Guo}, {Halverson}, {Harding}, {Harris}, {Hasselquist},
  {Hawley}, {Hayden}, {Hearty}, {Herrero Dav{\'o}}, {Ho}, {Hogg}, {Holtzman},
  {Honscheid}, {Huehnerhoff}, {Ivans}, {Jackson}, {Jiang}, {Johnson},
  {Kinemuchi}, {Kirkby}, {Klaene}, {Kneib}, {Koesterke}, {Lan}, {Lang}, {Le
  Goff}, {Leauthaud}, {Lee}, {Lee}, {Long}, {Loomis}, {Lucatello}, {Lupton},
  {Ma}, {Mack}, {Mahadevan}, {Maia}, {Majewski}, {Malanushenko},
  {Malanushenko}, {Manchado}, {Manera}, {Maraston}, {Margala}, {Martell},
  {Masters}, {McBride}, {McGreer}, {McMahon}, {M{\'e}nard}, {M{\'e}sz{\'a}ros},
  {Miralda-Escud{\'e}}, {Miyatake}, {Montero-Dorta}, {Montesano}, {More},
  {Morrison}, {Muna}, {Munn}, {Myers}, {Nguyen}, {Nichol}, {Nidever},
  {Noterdaeme}, {Nuza}, {O'Connell}, {O'Connell}, {O'Connell}, {Olmstead},
  {Oravetz}, {Owen}, {Padmanabhan}, {Palanque-Delabrouille}, {Pan}, {Parejko},
  {Parihar}, {P{\^a}ris}, {Pepper}, {Percival}, {P{\'e}rez-R{\`a}fols}, {Dotto
  Perottoni}, {Petitjean}, {Pieri}, {Pinsonneault}, {Prada}, {Price-Whelan},
  {Raddick}, {Rahman}, {Rebolo}, {Reid}, {Richards}, {Riffel}, {Robin},
  {Rocha-Pinto}, {Rockosi}, {Roe}, {Ross}, {Ross}, {Rossi}, {Roy},
  {Rubi{\~n}o-Martin}, {Sabiu}, {S{\'a}nchez}, {Santiago}, {Sayres},
  {Schiavon}, {Schlegel}, {Schlesinger}, {Schmidt}, {Schneider}, {Schultheis},
  {Sellgren}, {Seo}, {Shen}, {Shetrone}, {Shu}, {Simmons}, {Skrutskie},
  {Slosar}, {Smith}, {Snedden}, {Sobeck}, {Sobreira}, {Stassun}, {Steinmetz},
  {Strauss}, {Streblyanska}, {Suzuki}, {Swanson}, {Terrien}, {Thakar},
  {Thomas}, {Thompson}, {Tinker}, {Tojeiro}, {Troup}, {Vandenberg}, {Vargas
  Maga{\~n}a}, {Viel}, {Vogt}, {Wake}, {Weaver}, {Weinberg}, {Weiner}, {White},
  {White}, {Wilson}, {Wisniewski}, {Wood-Vasey}, {Y{\`e}che}, {York}, {Zamora},
  {Zasowski}, {Zehavi}, {Zhao}, {Zheng}, \& {Zhu}}]{ahn14}
{Ahn}, C.~P., {Alexandroff}, R., {Allende Prieto}, C., {et~al.} 2014, \apjs,
  211, 17

\bibitem[{{Andrews} \& {Martini}(2013)}]{Andrews_2013}
{Andrews}, B.~H. \& {Martini}, P. 2013, \apj, 765, 140

\bibitem[{{Andrews} \& {Smith}(2018)}]{AS17}
{Andrews}, J.~E. \& {Smith}, N. 2018, \mnras, 477, 74

\bibitem[{{Arcavi} {et~al.}(2017){Arcavi}, {Howell}, {Kasen}, {Bildsten},
  {Hosseinzadeh}, {McCully}, {Wong}, {Katz}, {Gal-Yam}, {Sollerman}, {Taddia},
  {Leloudas}, {Fremling}, {Nugent}, {Horesh}, {Mooley}, {Rumsey}, {Cenko},
  {Graham}, {Perley}, {Nakar}, {Shaviv}, {Bromberg}, {Shen}, {Ofek}, {Cao},
  {Wang}, {Huang}, {Rui}, {Zhang}, {Li}, {Li}, {Zhang}, {Valenti}, {Guevel},
  {Shappee}, {Kochanek}, {Holoien}, {Filippenko}, {Fender}, {Nyholm}, {Yaron},
  {Kasliwal}, {Sullivan}, {Blagorodnova}, {Walters}, {Lunnan}, {Khazov},
  {Andreoni}, {Laher}, {Konidaris}, {Wozniak}, \& {Bue}}]{Arcavi17}
{Arcavi}, I., {Howell}, D.~A., {Kasen}, D., {et~al.} 2017, \nat, 551, 210

\bibitem[{{Asplund} {et~al.}(2009){Asplund}, {Grevesse}, {Sauval}, \&
  {Scott}}]{Asplund_2009}
{Asplund}, M., {Grevesse}, N., {Sauval}, A.~J., \& {Scott}, P. 2009, \araa, 47,
  481

\bibitem[{{Bellm} {et~al.}(2019){Bellm}, {Kulkarni}, {Graham}, {Dekany},
  {Smith}, {Riddle}, {Masci}, {Helou}, {Prince}, {Adams}, {Barbarino},
  {Barlow}, {Bauer}, {Beck}, {Belicki}, {Biswas}, {Blagorodnova}, {Bodewits},
  {Bolin}, {Brinnel}, {Brooke}, {Bue}, {Bulla}, {Burruss}, {Cenko}, {Chang},
  {Connolly}, {Coughlin}, {Cromer}, {Cunningham}, {De}, {Delacroix}, {Desai},
  {Duev}, {Eadie}, {Farnham}, {Feeney}, {Feindt}, {Flynn}, {Franckowiak},
  {Frederick}, {Fremling}, {Gal-Yam}, {Gezari}, {Giomi}, {Goldstein},
  {Golkhou}, {Goobar}, {Groom}, {Hacopians}, {Hale}, {Henning}, {Ho}, {Hover},
  {Howell}, {Hung}, {Huppenkothen}, {Imel}, {Ip}, {Ivezi{\'c}}, {Jackson},
  {Jones}, {Juric}, {Kasliwal}, {Kaspi}, {Kaye}, {Kelley}, {Kowalski},
  {Kramer}, {Kupfer}, {Landry}, {Laher}, {Lee}, {Lin}, {Lin}, {Lunnan},
  {Giomi}, {Mahabal}, {Mao}, {Miller}, {Monkewitz}, {Murphy}, {Ngeow},
  {Nordin}, {Nugent}, {Ofek}, {Patterson}, {Penprase}, {Porter}, {Rauch},
  {Rebbapragada}, {Reiley}, {Rigault}, {Rodriguez}, {van Roestel}, {Rusholme},
  {van Santen}, {Schulze}, {Shupe}, {Singer}, {Soumagnac}, {Stein}, {Surace},
  {Sollerman}, {Szkody}, {Taddia}, {Terek}, {Van Sistine}, {van Velzen},
  {Vestrand}, {Walters}, {Ward}, {Ye}, {Yu}, {Yan}, \& {Zolkower}}]{Bellm_2018}
{Bellm}, E.~C., {Kulkarni}, S.~R., {Graham}, M.~J., {et~al.} 2019, \pasp, 131,
  018002

\bibitem[{{Blagorodnova} {et~al.}(2018){Blagorodnova}, {Neill}, {Walters},
  {Kulkarni}, {Fremling}, {Ben-Ami}, {Dekany}, {Fucik}, {Konidaris}, {Nash},
  {Ngeow}, {Ofek}, {O' Sullivan}, {Quimby}, {Ritter}, \& {Vyhmeister}}]{SEDM}
{Blagorodnova}, N., {Neill}, J.~D., {Walters}, R., {et~al.} 2018, \pasp, 130,
  035003

\bibitem[{{Burrows} {et~al.}(2005){Burrows}, {Hill}, {Nousek}, {Kennea},
  {Wells}, {Osborne}, {Abbey}, {Beardmore}, {Mukerjee}, {Short}, {Chincarini},
  {Campana}, {Citterio}, {Moretti}, {Pagani}, {Tagliaferri}, {Giommi},
  {Capalbi}, {Tamburelli}, {Angelini}, {Cusumano}, {Br{\"a}uninger}, {Burkert},
  \& {Hartner}}]{Burrows2005}
{Burrows}, D.~N., {Hill}, J.~E., {Nousek}, J.~A., {et~al.} 2005, \ssr, 120, 165

\bibitem[{{Byler} {et~al.}(2017){Byler}, {Dalcanton}, {Conroy}, \&
  {Johnson}}]{Byler2017a}
{Byler}, N., {Dalcanton}, J.~J., {Conroy}, C., \& {Johnson}, B.~D. 2017, \apj,
  840, 44

\bibitem[{{Calzetti} {et~al.}(2000){Calzetti}, {Armus}, {Bohlin}, {Kinney},
  {Koornneef}, \& {Storchi-Bergmann}}]{Calzetti2000a}
{Calzetti}, D., {Armus}, L., {Bohlin}, R.~C., {et~al.} 2000, \apj, 533, 682

\bibitem[{{Cardelli} {et~al.}(1989){Cardelli}, {Clayton}, \&
  {Mathis}}]{1989ApJ...345..245C}
{Cardelli}, J.~A., {Clayton}, G.~C., \& {Mathis}, J.~S. 1989, \apj, 345, 245

\bibitem[{{Cenko} {et~al.}(2006){Cenko}, {Fox}, {Moon}, {Harrison}, {Kulkarni},
  {Henning}, {Guzman}, {Bonati}, {Smith}, {Thicksten}, {Doyle}, {Petrie},
  {Gal-Yam}, {Soderberg}, {Anagnostou}, \& {Laity}}]{2006PASP..118.1396C}
{Cenko}, S.~B., {Fox}, D.~B., {Moon}, D.-S., {et~al.} 2006, \pasp, 118, 1396

\bibitem[{{Chabrier}(2003)}]{Chabrier2003a}
{Chabrier}, G. 2003, \pasp, 115, 763

\bibitem[{{Chambers} {et~al.}(2016){Chambers}, {Magnier}, {Metcalfe},
  {Flewelling}, {Huber}, {Waters}, {Denneau}, {Draper}, {Farrow}, {Finkbeiner},
  {Holmberg}, {Koppenhoefer}, {Price}, {Rest}, {Saglia}, {Schlafly}, {Smartt},
  {Sweeney}, {Wainscoat}, {Burgett}, {Chastel}, {Grav}, {Heasley}, {Hodapp},
  {Jedicke}, {Kaiser}, {Kudritzki}, {Luppino}, {Lupton}, {Monet}, {1Morgan},
  {Onaka}, {Shiao}, {Stubbs}, {Tonry}, {White}, {Ba{\~n}ados}, {Bell},
  {Bender}, {Bernard}, {Boegner}, {Boffi}, {Botticella}, {Calamida},
  {Casertano}, {Chen}, {Chen}, {Cole}, {Deacon}, {Frenk}, {Fitzsimmons},
  {Gezari}, {Gibbs}, {Goessl}, {Goggia}, {Gourgue}, {Goldman}, {Grant},
  {Grebel}, {Hambly}, {Hasinger}, {Heavens}, {Heckman}, {Henderson}, {Henning},
  {Holman}, {Hopp}, {Ip}, {Isani}, {Jackson}, {Keyes}, {Koekemoer}, {Kotak},
  {Le}, {Liska}, {Long}, {Lucey}, {Liu}, {Martin}, {Masci}, {McLean}, {Mindel},
  {Misra}, {Morganson}, {Murphy}, {Obaika}, {Narayan}, {Nieto-Santisteban},
  {Norberg}, {Peacock}, {Pier}, {Postman}, {Primak}, {Rae}, {Rai}, {Riess},
  {Riffeser}, {Rix}, {R{\"o}ser}, {Russel}, {Rutz}, {Schilbach}, {Schultz},
  {Scolnic}, {Strolger}, {Szalay}, {Seitz}, {Small}, {Smith}, {Soderblom},
  {Taylor}, {Thomson}, {Taylor}, {Thakar}, {Thiel}, {Thilker}, {Unger},
  {Urata}, {Valenti}, {Wagner}, {Walder}, {Walter}, {Watters}, {Werner},
  {Wood-Vasey}, \& {Wyse}}]{Chambers2016a}
{Chambers}, K.~C., {Magnier}, E.~A., {Metcalfe}, N., {et~al.} 2016, arXiv
  e-prints, arXiv:1612.05560

\bibitem[{{Chugai}(2018)}]{Chugai18}
{Chugai}, N.~N. 2018, Astronomy Letters, 44, 370

\bibitem[{{Conroy} {et~al.}(2009){Conroy}, {Gunn}, \& {White}}]{Conroy2009a}
{Conroy}, C., {Gunn}, J.~E., \& {White}, M. 2009, \apj, 699, 486

\bibitem[{{Cook} {et~al.}(2019){Cook}, {Kasliwal}, {Van Sistine}, {Kaplan},
  {Sutter}, {Kupfer}, {Shupe}, {Laher}, {Masci}, {Dale}, {Sesar}, {Brady},
  {Yan}, {Ofek}, {Reitze}, \& {Kulkarni}}]{CLU}
{Cook}, D.~O., {Kasliwal}, M.~M., {Van Sistine}, A., {et~al.} 2019, \apj, 880,
  7

\bibitem[{{Dekany} {et~al.}(2020){Dekany}, {Smith}, {Riddle}, {Feeney},
  {Porter}, {Hale}, {Zolkower}, {Belicki}, {Kaye}, {Henning}, {Walters},
  {Cromer}, {Delacroix}, {Rodriguez}, {Reiley}, {Mao}, {Hover}, {Murphy},
  {Burruss}, {Baker}, {Kowalski}, {Reif}, {Mueller}, {Bellm}, {Graham}, \&
  {Kulkarni}}]{dekany2020}
{Dekany}, R., {Smith}, R.~M., {Riddle}, R., {et~al.} 2020, \pasp, 132, 038001

\bibitem[{{Dessart}(2018)}]{Dessart18}
{Dessart}. 2018, \aap, 610, L10

\bibitem[{{Dopita} {et~al.}(2016){Dopita}, {Kewley}, {Sutherland}, \&
  {Nicholls}}]{Dopita_2016}
{Dopita}, M.~A., {Kewley}, L.~J., {Sutherland}, R.~S., \& {Nicholls}, D.~C.
  2016, \apss, 361, 61

\bibitem[{{Evans} {et~al.}(2009){Evans}, {Beardmore}, {Page}, {Osborne},
  {O'Brien}, {Willingale}, {Starling}, {Burrows}, {Godet}, {Vetere}, {Racusin},
  {Goad}, {Wiersema}, {Angelini}, {Capalbi}, {Chincarini}, {Gehrels}, {Kennea},
  {Margutti}, {Morris}, {Mountford}, {Pagani}, {Perri}, {Romano}, \&
  {Tanvir}}]{Evans2009a}
{Evans}, P.~A., {Beardmore}, A.~P., {Page}, K.~L., {et~al.} 2009, \mnras, 397,
  1177

\bibitem[{{Evans} {et~al.}(2007){Evans}, {Beardmore}, {Page}, {Tyler},
  {Osborne}, {Goad}, {O'Brien}, {Vetere}, {Racusin}, {Morris}, {Burrows},
  {Capalbi}, {Perri}, {Gehrels}, \& {Romano}}]{Evans2007a}
{Evans}, P.~A., {Beardmore}, A.~P., {Page}, K.~L., {et~al.} 2007, \aap, 469,
  379

\bibitem[{{Foreman-Mackey} {et~al.}(2014){Foreman-Mackey}, {Sick}, \&
  {Johnson}}]{ForemanMackey2014a}
{Foreman-Mackey}, D., {Sick}, J., \& {Johnson}, B. 2014, {Python-Fsps: Python
  Bindings To Fsps (V0.1.1)}

\bibitem[{{F{\"o}rster} {et~al.}(2020){F{\"o}rster}, {Cabrera-Vives},
  {Castillo-Navarrete}, {Est{\'e}vez}, {S{\'a}nchez-S{\'a}ez}, {Arredondo},
  {Bauer}, {Carrasco-Davis}, {Catelan}, {Elorrieta}, {Eyheramendy}, {Huijse},
  {Pignata}, {Reyes}, {Reyes}, {Rodr{\'\i}guez-Mancini}, {Ruz-Mieres},
  {Valenzuela}, {Alvarez-Maldonado}, {Astorga}, {Borissova}, {Clocchiatti}, {De
  Cicco}, {Donoso-Oliva}, {Graham}, {Kurtev}, {Mahabal}, {Maureira},
  {Molina-Ferreiro}, {Moya}, {Palma}, {P{\'e}rez-Carrasco}, {Protopapas},
  {Romero}, {Sabatini-Gacit{\'u}a}, {S{\'a}nchez}, {San Mart{\'\i}n},
  {Sep{\'u}lveda-Cobo}, {Vera}, \& {Vergara}}]{alerce}
{F{\"o}rster}, F., {Cabrera-Vives}, G., {Castillo-Navarrete}, E., {et~al.}
  2020, arXiv e-prints, arXiv:2008.03303

\bibitem[{{Fremling} {et~al.}(2016){Fremling}, {Sollerman}, {Taddia}, {Ergon},
  {Fraser}, {Karamehmetoglu}, {Valenti}, {Jerkstrand }, {Arcavi}, {Bufano},
  {Elias Rosa}, {Filippenko}, {Fox}, {Gal-Yam}, {Howell}, {Kotak}, {Mazzali},
  {Milisavljevic}, {Nugent}, {Nyholm}, {Pian}, \& {Smartt}}]{fremling16}
{Fremling}, C., {Sollerman}, J., {Taddia}, F., {et~al.} 2016, \aap, 593, A68

\bibitem[{{Gehrels} {et~al.}(2004){Gehrels}, {Chincarini}, {Giommi}, {Mason},
  {Nousek}, {Wells}, {White}, {Barthelmy}, {Burrows}, {Cominsky}, {Hurley},
  {Marshall}, {M{\'e}sz{\'a}ros}, {Roming}, {Angelini}, {Barbier}, {Belloni},
  {Campana}, {Caraveo}, {Chester}, {Citterio}, {Cline}, {Cropper}, {Cummings},
  {Dean}, {Feigelson}, {Fenimore}, {Frail}, {Fruchter}, {Garmire}, {Gendreau},
  {Ghisellini}, {Greiner}, {Hill}, {Hunsberger}, {Krimm}, {Kulkarni}, {Kumar},
  {Lebrun}, {Lloyd-Ronning}, {Markwardt}, {Mattson}, {Mushotzky}, {Norris},
  {Osborne}, {Paczynski}, {Palmer}, {Park}, {Parsons}, {Paul}, {Rees},
  {Reynolds}, {Rhoads}, {Sasseen}, {Schaefer}, {Short}, {Smale}, {Smith},
  {Stella}, {Tagliaferri}, {Takahashi}, {Tashiro}, {Townsley}, {Tueller},
  {Turner}, {Vietri}, {Voges}, {Ward}, {Willingale}, {Zerbi}, \&
  {Zhang}}]{gehrels2004}
{Gehrels}, N., {Chincarini}, G., {Giommi}, P., {et~al.} 2004, \apj, 611, 1005

\bibitem[{Graham {et~al.}(2019)Graham, Kulkarni, Bellm, Adams, Barbarino,
  Blagorodnova, Bodewits, Bolin, Brady, Cenko, \& et~al.}]{Graham_2019}
Graham, M.~J., Kulkarni, S.~R., Bellm, E.~C., {et~al.} 2019, \pasp, 131, 078001

\bibitem[{{HI4PI Collaboration} {et~al.}(2016){HI4PI Collaboration}, {Ben
  Bekhti}, {Fl{\"o}er}, {Keller}, {Kerp}, {Lenz}, {Winkel}, {Bailin},
  {Calabretta}, {Dedes}, {Ford}, {Gibson}, {Haud}, {Janowiecki}, {Kalberla},
  {Lockman}, {McClure-Griffiths}, {Murphy}, {Nakanishi}, {Pisano}, \&
  {Staveley-Smith}}]{HI4PI2016a}
{HI4PI Collaboration}, {Ben Bekhti}, N., {Fl{\"o}er}, L., {et~al.} 2016, \aap,
  594, A116

\bibitem[{{Howell} {et~al.}(2005){Howell}, {Sullivan}, {Perrett}, {Bronder},
  {Hook}, {Astier}, {Aubourg}, {Balam}, {Basa}, {Carlberg}, {Fabbro},
  {Fouchez}, {Guy}, {Lafoux}, {Neill}, {Pain}, {Palanque-Delabrouille},
  {Pritchet}, {Regnault}, {Rich}, {Taillet}, {Knop}, {McMahon}, {Perlmutter},
  \& {Walton}}]{Howell2005}
{Howell}, D.~A., {Sullivan}, M., {Perrett}, K., {et~al.} 2005, \apj, 634, 1190

\bibitem[{{Kasliwal} {et~al.}(2019){Kasliwal}, {Cannella}, {Bagdasaryan},
  {Hung}, {Feindt}, {Singer}, {Coughlin}, {Fremling}, {Walters}, {Duev},
  {Itoh}, \& {Quimby}}]{growth}
{Kasliwal}, M.~M., {Cannella}, C., {Bagdasaryan}, A., {et~al.} 2019, \pasp,
  131, 038003

\bibitem[{{Lang}(2014)}]{Lang2014a}
{Lang}, D. 2014, \aj, 147, 108

\bibitem[{{Leja} {et~al.}(2017){Leja}, {Johnson}, {Conroy}, {van Dokkum}, \&
  {Byler}}]{Leja2017a}
{Leja}, J., {Johnson}, B.~D., {Conroy}, C., {van Dokkum}, P.~G., \& {Byler}, N.
  2017, \apj, 837, 170

\bibitem[{{Mainzer} {et~al.}(2014){Mainzer}, {Bauer}, {Cutri}, {Grav},
  {Masiero}, {Beck}, {Clarkson}, {Conrow}, {Dailey}, {Eisenhardt}, {Fabinsky},
  {Fajardo-Acosta}, {Fowler}, {Gelino}, {Grillmair}, {Heinrichsen}, {Kendall},
  {Kirkpatrick}, {Liu}, {Masci}, {McCallon}, {Nugent}, {Papin}, {Rice},
  {Royer}, {Ryan}, {Sevilla}, {Sonnett}, {Stevenson}, {Thompson}, {Wheelock},
  {Wiemer}, {Wittman}, {Wright}, \& {Yan}}]{Mainzer2014a}
{Mainzer}, A., {Bauer}, J., {Cutri}, R.~M., {et~al.} 2014, \apj, 792, 30

\bibitem[{{Martin} {et~al.}(2005){Martin}, {Fanson}, {Schiminovich},
  {Morrissey}, {Friedman}, {Barlow}, {Conrow}, {Grange}, {Jelinsky},
  {Milliard}, {Siegmund}, {Bianchi}, {Byun}, {Donas}, {Forster}, {Heckman},
  {Lee}, {Madore}, {Malina}, {Neff}, {Rich}, {Small}, {Surber}, {Szalay},
  {Welsh}, \& {Wyder}}]{Martin2005a}
{Martin}, D.~C., {Fanson}, J., {Schiminovich}, D., {et~al.} 2005, \apj, 619, L1

\bibitem[{{Masci} {et~al.}(2019){Masci}, {Laher}, {Rusholme}, {Shupe}, {Groom},
  {Surace}, {Jackson}, {Monkewitz}, {Beck}, {Flynn}, {Terek}, {Landry},
  {Hacopians}, {Desai}, {Howell}, {Brooke}, {Imel}, {Wachter}, {Ye}, {Lin},
  {Cenko}, {Cunningham}, {Rebbapragada}, {Bue}, {Miller}, {Mahabal}, {Bellm},
  {Patterson}, {Juri{\'c}}, {Golkhou}, {Ofek}, {Walters}, {Graham}, {Kasliwal},
  {Dekany}, {Kupfer}, {Burdge}, {Cannella}, {Barlow}, {Van Sistine}, {Giomi},
  {Fremling}, {Blagorodnova}, {Levitan}, {Riddle}, {Smith}, {Helou}, {Prince},
  \& {Kulkarni}}]{2019PASP..131a8003M}
{Masci}, F.~J., {Laher}, R.~R., {Rusholme}, B., {et~al.} 2019, \pasp, 131,
  018003

\bibitem[{{Meisner} {et~al.}(2017){Meisner}, {Lang}, \&
  {Schlegel}}]{Meisner2017a}
{Meisner}, A.~M., {Lang}, D., \& {Schlegel}, D.~J. 2017, \aj, 153, 38

\bibitem[{Moriya {et~al.}(2019)Moriya, Mazzali, \& Pian}]{2020MNRAS.491.1384M}
Moriya, T.~J., Mazzali, P.~A., \& Pian, E. 2019, \mnras, 491, 1384

\bibitem[{Nadyozhin(2003)}]{Nadyozhin}
Nadyozhin, D.~K. 2003, \mnras, 346, 97

\bibitem[{{Perley} {et~al.}(2020){Perley}, {Taggart}, {Dahiwale}, \&
  {Fremling}}]{2020TNSCR.987....1P}
{Perley}, D.~A., {Taggart}, K., {Dahiwale}, A., \& {Fremling}, C. 2020,
  Transient Name Server Classification Report, 2020-987, 1

\bibitem[{{Pettini} \& {Pagel}(2004)}]{Pettini_2004}
{Pettini}, M. \& {Pagel}, B. E.~J. 2004, \mnras, 348, L59

\bibitem[{{Rigault} {et~al.}(2019){Rigault}, {Neill}, {Blagorodnova}, {Dugas},
  {Feeney}, {Walters}, {Brinnel}, {Copin}, {Fremling}, {Nordin}, \&
  {Sollerman}}]{RigaultPySEDM}
{Rigault}, M., {Neill}, J.~D., {Blagorodnova}, N., {et~al.} 2019, \aap, 627,
  A115

\bibitem[{{Roming} {et~al.}(2005){Roming}, {Kennedy}, {Mason}, {Nousek}, {Ahr},
  {Bingham}, {Broos}, {Carter}, {Hancock}, {Huckle}, {Hunsberger}, {Kawakami},
  {Killough}, {Koch}, {McLelland}, {Smith}, {Smith}, {Soto}, {Boyd},
  {Breeveld}, {Holland}, {Ivanushkina}, {Pryzby}, {Still}, \&
  {Stock}}]{2005SSRv..120...95R}
{Roming}, P. W.~A., {Kennedy}, T.~E., {Mason}, K.~O., {et~al.} 2005, \ssr, 120,
  95

\bibitem[{{Schulze} {et~al.}(2020){Schulze}, {Yaron}, {Sollerman}, {Leloudas},
  {Gal}, {Wright}, {Lunnan}, {Gal-Yam}, {Ofek}, {Perley}, {Filippenko},
  {Kasliwal}, {Kulkarni}, {Nugent}, {Quimby}, {Sullivan}, {Linn Strothjohann},
  {Arcavi}, {Ben-Ami}, {Bianco}, {Bloom}, {De}, {Fraser}, {Fremling}, {Horesh},
  {Johansson}, {Kelly}, {Knezevic}, {Maguire}, {Nyholm}, {Semeli
  Papadogiannakis}, {Petrushevska}, {Rubin}, {Yan}, {Yang}, {Adams}, {Bufano},
  {Clubb}, {Foley}, {Green}, {Harmanen}, {Ho}, {Hook}, {Hosseinzadeh},
  {Howell}, {Kong}, {Kotak}, {Matheson}, {McCully}, {Milisavljevic}, {Pan},
  {Poznanski}, {Shivvers}, \& {van Velzen}}]{Schulze2020a}
{Schulze}, S., {Yaron}, O., {Sollerman}, J., {et~al.} 2020, arXiv e-prints,
  arXiv:2008.05988

\bibitem[{{Skrutskie} {et~al.}(2006){Skrutskie}, {Cutri}, {Stiening},
  {Weinberg}, {Schneider}, {Carpenter}, {Beichman}, {Capps}, {Chester},
  {Elias}, {Huchra}, {Liebert}, {Lonsdale}, {Monet}, {Price}, {Seitzer},
  {Jarrett}, {Kirkpatrick}, {Gizis}, {Howard}, {Evans}, {Fowler}, {Fullmer},
  {Hurt}, {Light}, {Kopan}, {Marsh}, {McCallon}, {Tam}, {Van Dyk}, \&
  {Wheelock}}]{Skrutskie2006a}
{Skrutskie}, M.~F., {Cutri}, R.~M., {Stiening}, R., {et~al.} 2006, \aj, 131,
  1163

\bibitem[{Soker \& Gilkis(2017)}]{Soker18}
Soker, N. \& Gilkis, A. 2017, \mnras, 475, 1198

\bibitem[{{Sollerman} {et~al.}(2020){Sollerman}, {Fransson}, {Barbarino},
  {Fremling}, {Horesh}, {Kool}, {Schulze}, {Sfaradi}, {Yang}, {Bellm},
  {Burruss}, {Cunningham}, {De}, {Drake}, {Golkhou}, {Green}, {Kasliwal},
  {Kulkarni}, {Kupfer}, {Laher}, {Masci}, {Rodriguez}, {Rusholme}, {Williams},
  {Yan}, \& {Zolkower}}]{Sollerman2020}
{Sollerman}, J., {Fransson}, C., {Barbarino}, C., {et~al.} 2020, arXiv
  e-prints, arXiv:2009.04154

\bibitem[{{Sollerman} {et~al.}(2019){Sollerman}, {Taddia}, {Arcavi},
  {Fremling}, {Fransson}, {Burke}, {Cenko}, {Andersen}, {Andreoni},
  {Barbarino}, {Blagorodova}, {Brink}, {Filippenko}, {Gal-Yam}, {Hiramatsu},
  {Hosseinzadeh}, {Howell}, {de Jaeger}, {Lunnan}, {McCully}, {Perley},
  {Tartaglia}, {Terreran}, {Valenti}, \& {Wang}}]{Sollerman2019}
{Sollerman}, J., {Taddia}, F., {Arcavi}, I., {et~al.} 2019, \aap, 621, A30

\bibitem[{{Speagle}(2020)}]{Speagle2020a}
{Speagle}, J.~S. 2020, \mnras, 493, 3132

\bibitem[{{Terreran} {et~al.}(2017){Terreran}, {Pumo}, {Chen}, {Moriya},
  {Taddia}, {Dessart}, {Zampieri}, {Smartt}, {Benetti}, {Inserra},
  {Cappellaro}, {Nicholl}, {Fraser}, {Wyrzykowski}, {Udalski}, {Howell},
  {McCully}, {Valenti}, {Dimitriadis}, {Maguire}, {Sullivan}, {Smith}, {Yaron},
  {Young}, {Anderson}, {Della Valle}, {Elias-Rosa}, {Gal-Yam}, {Jerkstrand},
  {Kankare}, {Pastorello}, {Sollerman}, {Turatto}, {Kostrzewa-Rutkowska},
  {Koz{\l}owski}, {Mr{\'o}z}, {Pawlak}, {Pietrukowicz}, {Poleski}, {Skowron},
  {Skowron}, {Soszy{\'n}ski}, {Szyma{\'n}ski}, \& {Ulaczyk}}]{Terreran}
{Terreran}, G., {Pumo}, M.~L., {Chen}, T.~W., {et~al.} 2017, Nature Astronomy,
  1, 713

\bibitem[{{Tonry} {et~al.}(2020){Tonry}, {Denneau}, {Heinze}, {Weiland},
  {Flewelling}, {Stalder}, {Rest}, {Stubbs}, {Smith}, {Smartt}, {Young},
  {Srivastav}, {McBrien}, {O'Neill}, {Clark}, {Fulton}, {Gillanders}, {Dobson},
  {Chen}, {Wright}, \& {Anderson}}]{2020TNSTR.869....1T}
{Tonry}, J., {Denneau}, L., {Heinze}, A., {et~al.} 2020, Transient Name Server
  Discovery Report, 2020-869, 1

\bibitem[{Valenti {et~al.}(2013)Valenti, Sand, Pastorello, Graham, Howell,
  Parrent, Tomasella, Ochner, Fraser, Benetti, Yuan, Smartt, Maund, Arcavi,
  Gal-Yam, Inserra, \& Young}]{Valenti_2013}
Valenti, S., Sand, D., Pastorello, A., {et~al.} 2013, \mnras, 438, L101

\bibitem[{{Villar} {et~al.}(2019){Villar}, {Berger}, {Miller}, {Chornock},
  {Rest}, {Jones}, {Drout}, {Foley}, {Kirshner}, {Lunnan}, {Magnier},
  {Milisavljevic}, {Sanders}, \& {Scolnic}}]{villar}
{Villar}, V.~A., {Berger}, E., {Miller}, G., {et~al.} 2019, \apj, 884, 83

\bibitem[{{Wang} {et~al.}(2018){Wang}, {Wang}, {Wang}, {Dai}, {Liu}, {Song},
  {Rui}, {Cano}, \& {Li}}]{WangWangWang18}
{Wang}, L.~J., {Wang}, X.~F., {Wang}, S.~Q., {et~al.} 2018, \apj, 865, 95

\bibitem[{{Woosley}(2018)}]{Woosley18}
{Woosley}, S.~E. 2018, \apj, 863, 105

\bibitem[{{Wright} {et~al.}(2016){Wright}, {Robotham}, {Bourne}, {Driver},
  {Dunne}, {Maddox}, {Alpaslan}, {Andrews}, {Bauer}, {Bland-Hawthorn},
  {Brough}, {Brown}, {Clarke}, {Cluver}, {Davies}, {Grootes}, {Holwerda},
  {Hopkins}, {Jarrett}, {Kafle}, {Lange}, {Liske}, {Loveday}, {Moffett},
  {Norberg}, {Popescu}, {Smith}, {Taylor}, {Tuffs}, {Wang}, \&
  {Wilkins}}]{Wright2016a}
{Wright}, A.~H., {Robotham}, A.~S.~G., {Bourne}, N., {et~al.} 2016, \mnras,
  460, 765

\bibitem[{{Wright} {et~al.}(2010){Wright}, {Eisenhardt}, {Mainzer}, {Ressler},
  {Cutri}, {Jarrett}, {Kirkpatrick}, {Padgett}, {McMillan}, {Skrutskie},
  {Stanford}, {Cohen}, {Walker}, {Mather}, {Leisawitz}, {Gautier}, {McLean},
  {Benford}, {Lonsdale}, {Blain}, {Mendez}, {Irace}, {Duval}, {Liu}, {Royer},
  {Heinrichsen}, {Howard}, {Shannon}, {Kendall}, {Walsh}, {Larsen}, {Cardon},
  {Schick}, {Schwalm}, {Abid}, {Fabinsky}, {Naes}, \& {Tsai}}]{Wright2010a}
{Wright}, E.~L., {Eisenhardt}, P.~R.~M., {Mainzer}, A.~K., {et~al.} 2010, \aj,
  140, 1868

\bibitem[{{Yaron} \& {Gal-Yam}(2012)}]{wiserep}
{Yaron}, O. \& {Gal-Yam}, A. 2012, \pasp, 124, 668

\end{thebibliography}

\onecolumn

\begin{deluxetable}{lccc}
\tablewidth{0pt}
\tabletypesize{\scriptsize}
\tablecaption{Summary of Spectroscopic Observations\label{tab:spec}}
\tablehead{
\colhead{Object} &
\colhead{Observation Date} & 
\colhead{Phase} &
\colhead{Telescope+Instrument} \\
\colhead{} &
\colhead{(YYYY MM DD)}  & 
\colhead{(Rest-frame days)} &
\colhead{} 
}
\startdata
SN\,2020faa & 2020 Mar 31 & 6.4 & P60+SEDM \\
SN\,2020faa & 2020 Apr 05 & 11.9 & LT+SPRAT \\
SN\,2020faa & 2020 Jun 01 & 66.1 & P60+SEDM \\
SN\,2020faa & 2020 Jun 21 & 85.2 &P60+SEDM \\
SN\,2020faa & 2020 Jul 02 & 96.4 & NOT+ALFOSC \\
SN\,2020faa & 2020 Jul 24 & 117.6 & NOT+ALFOSC \\
SN\,2020faa & 2020 Jul 26 & 118.8 & P60+SEDM \\
SN\,2020faa & 2020 Aug 01 & 124.5 & P60+SEDM \\
SN\,2020faa & 2020 Aug 11 & 134.1 & P60+SEDM \\
SN\,2020faa & 2020 Aug 15 & 138.6 & NOT+ALFOSC \\
SN\,2020faa & 2020 Aug 21 & 143.7 & P60+SEDM \\  
SN\,2020faa & 2020 Aug 24 & 146.6 & P60+SEDM \\
SN\,2020faa & 2020 Aug 30 & 153.0 & NOT+ALFOSC \\
SN\,2020faa & 2020 Sep 06 & 159.0 & P60+SEDM \\
\enddata
\end{deluxetable}

\begin{deluxetable}{lccc}
\tablewidth{0pt}
\tabletypesize{\scriptsize}
\tablecaption{Host galaxy photometry\label{tab:host}}
\tablehead{
\colhead{Survey} &
\colhead{Filter} &
\colhead{Wavelength} & 
\colhead{Brightness} \\
\colhead{} &
\colhead{} &
\colhead{(\AA)} & 
\colhead{(AB mag)} 
}
\startdata
GALEX   &$ FUV   $& 1549.0      &$ 19.30 \pm 0.16$\\
GALEX   &$ NUV   $& 2304.7      &$ 18.68 \pm 0.07$\\
PS1     &$ g     $& 4810.9      &$ 17.10 \pm 0.03$\\
PS1     &$ r     $& 6156.4      &$ 16.79 \pm 0.03$\\
PS1     &$ i     $& 7503.7      &$ 16.56 \pm 0.03$\\
PS1     &$ z     $& 8668.6      &$ 16.51 \pm 0.03$\\
PS1     &$ y     $& 9613.5      &$ 16.39 \pm 0.06$\\
2MASS   &$ J     $& 12350       &$ 16.69 \pm 0.19$\\
2MASS   &$ H     $& 16620       &$ 16.23 \pm 0.24$\\
2MASS   &$ K     $& 21590       &$ 16.44 \pm 0.27$\\
\wise   &$ W1    $& 33526       &$ 16.83 \pm 0.04$\\
\wise   &$ W2    $& 46028       &$ 17.36 \pm 0.04$\\
\enddata
\tablecomments{Magnitudes are not corrected for extinction. 
The effective wavelengths of the filter response functions were taken from \href{http://svo2.cab.inta-csic.es/theory/fps/}{http://svo2.cab.inta-csic.es/theory/fps/}.}
\end{deluxetable}

\end{document}